\documentclass[review]{elsarticle}

\usepackage{tikz}
\usepackage{verbatim}
\usepackage{textcomp}
\usepackage{algpseudocode}
\usepackage{mdframed}
\usepackage[acronym]{glossaries}
\usepackage[normalem]{ulem}
\usepackage{epstopdf}
\usepackage{subcaption}
\usepackage{graphicx}
\usepackage{hyperref}
\usepackage{tikzpagenodes}

\usetikzlibrary{backgrounds, calc,trees,positioning,arrows,chains,shapes.geometric, decorations.pathreplacing,decorations.pathmorphing,shapes,patterns, matrix,shapes.symbols,fit,scopes,external}

\usepackage{pgfplots,pgfplotstable}
\pgfplotsset{compat=newest}

\usepackage{booktabs} 
\usepackage{graphicx}
\usepackage{balance}
\usepackage{multirow}
\usepackage[acronym]{glossaries}
\usepackage{listings}
\usepackage{xcolor} 
\usepackage{soul}
\usepackage{lineno,hyperref}
\modulolinenumbers[5]

\journal{Simulation Modelling Practice and Theory}

\newcommand{\cl}[1]{\textsc{#1}}
\newcommand{\attr}[1]{\textsc{#1}}
\newcommand{\kdos}[0]{$\mathsf{K2}$}

\newcommand\packet[3]{%
\begin{scope}[shift={(#1,#2)}]%
     \draw[#3] (0,0) rectangle (1,1);%
\end{scope}%
}

\definecolor{seablue}{RGB}{05,102,141}
\definecolor{alabamacrimson}{RGB}{172,00,54}
\definecolor{darkgreen}{RGB}{03,49,46}
\definecolor{metallicsunburst}{RGB}{168,118,62}
\definecolor{silver}{RGB}{193,193,193}

\long\def\/*#1*/{}

\begin{document}
\begin{frontmatter}

\title{An E2E Simulator for 5G NR Networks}

\author{Natale Patriciello}
\ead{natale.patriciello@cttc.cat}

\author{Sandra Lagen}
\ead{sandra.lagen@cttc.cat}

\author{Biljana Bojovic}
\ead{biljana.bojovic@cttc.cat}

\author{Lorenza Giupponi}
\ead{lorenza.giupponi@cttc.cat}

\address{Centre Tecnol\`ogic de Telecomunicacions de Catalunya (CTTC/CERCA) \\
Avinguda Carl Friedrich Gauss, 7 \\
08860 Castelldefels, Barcelona, Spain \\
\{npatriciello, slagen, bbojovic, lgiupponi\}@cttc.cat}


\begin{abstract}
	As the specification of the new 5G NR standard proceeds inside 3GPP, the availability of a versatile, full-stack, End-to-End (E2E), and open source simulator becomes a necessity to extract insights from the recently approved 3GPP specifications. This paper presents an extension to ns-3, a well-known discrete-event network simulator, to support the NR Radio Access Network. The present work describes the design and implementation choices at the MAC and PHY layers, and it discusses a technical solution for managing different bandwidth parts. Finally, we present calibration results, according to 3GPP procedures, and we show how to get E2E performance indicators in a realistic deployment scenario, with special emphasis on the E2E latency. 
\end{abstract}

\end{frontmatter}

\newacronym{scs}{SCS}{SubCarrier Spacing}
\newacronym{ofdm}{OFDM}{Orthogonal Frequency Division Multiplexing}
\newacronym{fdm}{FDM}{Frequency Division Multiplexing}
\newacronym{cpu}{CPU}{Central Processing Unit}
\newacronym{tcp}{TCP}{Transmission Control Protocol}
\newacronym{tcpw}{TCPW}{TCP Wave}
\newacronym{lte}{LTE}{Long Term Evolution}
\newacronym{nr}{NR}{New Radio}
\newacronym{cca}{CCA}{Congestion Control Algorithm}
\newacronym{cwnd}{cWnd}{Congestion Window}
\newacronym{caa}{CAA}{Congestion Avoidance Algorithm}
\newacronym{bbr}{BBR}{Bottleneck Bandwidth and Round-trip propagation time}
\newacronym{nv}{NV}{New Vegas}
\newacronym{rtt}{RTT}{Round-Trip Time}
\newacronym{ietf}{IETF}{Internet Engineering Task Force}
\newacronym{rfc}{RFC}{Request For Comments}
\newacronym{gcc}{GCC}{GNU Compiler Collection}
\newacronym{tso}{TSO}{TCP Segmentation Offloading}
\newacronym{tsq}{TSQ}{TCP Small Queue}
\newacronym{gbr}{GBR}{Guaranteed Bit Rate}
\newacronym{nongbr}{non-GBR}{non-Guaranteed Bit Rate}
\newacronym{enb}{eNB}{Evolved Node B}
\newacronym{dpi}{DPI}{Deep Packet Inspection}
\newacronym{rlc}{RLC}{Radio Link Control}
\newacronym{bsr}{BSR}{Buffer Status Report}
\newacronym{qos}{QoS}{Quality of Service}
\newacronym{aqm}{AQM}{Active Queue Management}
\newacronym{rds}{RDS}{Radio Data Scheduler}
\newacronym{tc}{TC}{Traffic Control}
\newacronym{drb}{DRB}{Data Radio Bearer}
\newacronym{rnti}{RNTI}{Radio Network Temporary Identifier}
\newacronym{bql}{BQL}{Byte Queue Limits}
\newacronym{ue}{UE}{User Equipment}
\newacronym{am}{AM}{Acknowledged Mode}
\newacronym{epc}{EPC}{Evolved Packet Core}
\newacronym{cn}{CN}{Core Network}
\newacronym{gnb}{gNB}{next-Generation Node B}
\newacronym{ran}{RAN}{Radio Access Network}
\newacronym{3gpp}{3GPP}{3rd Generation Partnership Project}
\newacronym{5g}{5G}{fifth Generation}
\newacronym{dl}{DL}{DownLink}
\newacronym{ul}{UL}{UpLink}
\newacronym{tti}{TTI}{Transmission Time Interval}
\newacronym{sr}{SR}{Scheduling Request}
\newacronym{e2e}{E2E}{End-To-End}
\newacronym{embb}{eMBB}{enhanced Mobile BroadBand}
\newacronym{urllc}{URLLC}{Ultra-Reliable and Low-Latency Communications}
\newacronym{mmtc}{mMTC}{massive Machine Type Communications}
\newacronym{phy}{PHY}{Physical}
\newacronym{mac}{MAC}{Medium Access Control}
\newacronym{prb}{PRB}{Physical Resource Block}
\newacronym{pps}{pps}{Packets Per Second}
\newacronym{cp}{CP}{Cyclic Prefix}
\newacronym{tbs}{TBS}{Transport Block Size}
\newacronym{tb}{TB}{Transport Block}
\newacronym{cb}{CB}{Code Block}
\newacronym{gtp}{GTP}{GPRS Tunneling Protocol}
\newacronym{sap}{SAP}{Service Access Point}
\newacronym{tm}{TM}{Transparent Mode}
\newacronym{um}{UM}{Unacknowledged Mode}
\newacronym{sm}{SM}{Saturation Mode}
\newacronym{rrc}{RRC}{Radio Resource Control}
\newacronym{tdma}{TDMA}{Time-Division Multiple Access}
\newacronym{ofdma}{OFDMA}{Orthogonal Frequency-Division Multiple Access}
\newacronym{rbg}{RBG}{Resource Block Group}
\newacronym{rb}{RB}{Resource Block}
\newacronym{dci}{DCI}{Downlink Control Information}
\newacronym{uci}{UCI}{Uplink Control Information}
\newacronym{ipat}{IPAT}{Inter-Packet Arrival Time}
\newacronym{pdsch}{PDSCH}{Physical Downlink Shared Channel}
\newacronym{pusch}{PUSCH}{Physical Uplink Shared Channel}
\newacronym{pucch}{PUCCH}{Physical Uplink Control Channel}
\newacronym{pdcch}{PDCCH}{Physical Downlink Control Channel}
\newacronym{tdd}{TDD}{Time Division Duplex}
\newacronym{fdd}{FDD}{Frequency Division Duplex}
\newacronym{rach}{RACH}{Random Access Channel}
\newacronym{cbr}{CBR}{Constant Bit Rate}
\newacronym{los}{LoS}{Line-of-Sight}
\newacronym{mcs}{MCS}{Modulation Coding Scheme}
\newacronym{bwp}{BWP}{Bandwidth Part}
\newacronym{cqi}{CQI}{Channel Quality Indicator}
\newacronym{bler}{BLER}{Block Error Rate}
\newacronym{tbler}{TBLER}{Transport Block Error Rate}
\newacronym{mi}{MI}{Mutual Information}
\newacronym{l2sm}{L2SM}{Link to System Mapping}
\newacronym{sliv}{SLIV}{Start and Length Indicator Value}
\newacronym{mmwave}{mmWave}{millimeter-wave}
\newacronym{pdu}{PDU}{Packet Data Unit}
\newacronym{ca}{CA}{Carrier Aggregation}
\newacronym{snr}{SNR}{Signal-to-Noise Ratio}
\newacronym{sinr}{SINR}{Signal to Interference-plus-Noise Ratio}
\newacronym{pdcp}{PDCP}{Packet Data Convergence Protocol}
\newacronym{sdap}{SDAP}{Service Data Adaptation Protocol}
\newacronym{sdu}{SDU}{Service Data Unit}
\newacronym{nas}{NAS}{Non-Access Stratum}
\newacronym{sme}{SME}{Small and Medium Enterprise}
\newacronym{rat}{RAT}{Radio Access Technology}
\newacronym{pgw}{PGW}{Packet data network GateWay}
\newacronym{sgw}{SGW}{Service GateWay}
\newacronym{ldpc}{LDPC}{low Density Parity Check}

\section{Introduction}
%
%
%
%

The \gls{3gpp} is devoting significant efforts to define the \gls{5g} \gls{nr} access technology~\cite{TS38300}, which has flexible, scalable, and forward-compatible \gls{phy} and \gls{mac} layers to support a wide range of center carrier frequencies, deployment options, and variety of use cases.
To account for that, and as compared to \gls{lte}, \gls{nr} includes new features, such as a flexible frame structure by means of multiple numerologies support, dynamic \gls{tdd}, support for new \gls{mmwave} frequency bands, beam management-related operations, support for wide channel bandwidth operations and frequency-division multiplexing of multiple bandwidth parts, symbol-level scheduling through mini-slots and variable \gls{tti}s, and new channel coding schemes. The new \gls{nr} features span over all the protocol stack, also introducing a new layer above \gls{pdcp}, called \gls{sdap}, standalone and non-standalone architectures, and a mandatory split of control and user planes in the core network.

Research institutions or \gls{sme}s that cannot develop sophisticated simulation tools capable of simulating 5G and beyond networks due to the cost, time effort, and required human resources, are at risk of being cut out from the early stages of the development process. Some of them rely on analytic simulation methods. However, the assumptions and simplifications in the sender and receiver nodes, as well as in other network segments and layers, limit the generality of the extracted results. Moreover, it is tough to represent external network dynamics (like the burstiness of data traffic) or to assess the interaction with the core network and the mobility of the users, without a solid full-stack \gls{e2e} simulation model. 

As researchers, we are not only interested in the low-level characterization of the previously mentioned \gls{nr} features but also we want to have an overall view of the system, which starts from the application level to the \gls{phy} layer and includes an \gls{e2e} performance evaluation from the \gls{ue} to the remote host. Our objective is to properly evaluate the performance of a sophisticated and flexible technology, \gls{nr}, and to be able to conduct interoperability studies with other technologies. As a result, in this work, we present an NR network simulator that has been built as a pluggable module to ns-3\footnote{www.nsnam.org}. The simulator models the \gls{nr} technology with a high-fidelity full protocol stack, and it has been calibrated according to 3GPP procedures. In particular, our simulator offers an abstraction of the \gls{phy} layer and high-fidelity implementations from the \gls{mac} to the application layer. It can be used to evaluate cross-layer and \gls{e2e} performance, as well as a platform to assess the coexistence of NR with other technologies. As an example of the capabilities offered by the simulator, we already used it to evaluate the impact of the processing and decoding times in the \gls{e2e} delay in~\cite{natale:18}, for different \gls{nr} numerologies. 

We believe that the open distribution of this simulator, under the terms of the GPLv2 license, represents an unprecedented contribution to the community and facilitates innovation in the area of 5G. The GPLv2 adoption is necessary because our simulator is derived from ns-3, which is GPLv2-licensed. Even if in a lot of GPL-covered software there is an explicit clause that permits the user to choose any later version of the license, the ns-3 simulator explicitly disabled this option, thus making our simulator effectively released under the sole GPLv2 license (as much as the Linux kernel). People used to GPLv3 software have to keep in mind the following main differences between the two GPL versions that apply to our software:
\begin{itemize}
  \item GPLv2 code cannot be combined with a range of software licenses that are now compatible with GPLv3 (the most prominent case is Apache v2, the license of the OpenAir Interface software);
  \item The GPLv2 license does not contain an explicit patent license.
\end{itemize}
There is a vast literature on the GPLv3 versus GPLv2 license, so for space constraints, we will not expand on all the specific differences. The interested reader is referred to the GPL
FAQ\footnote{https://www.gnu.org/licenses/gpl-faq.html} or the many resources available on the
Web\footnote{https://en.wikipedia.org/wiki/GNU\_General\_Public\_License}. We would like though to remind that, in the pure open spirit, we are using a very widely used license to foster the development and research around 5G concepts.

The objective of this paper is to give the reader a complete overview of the NR simulator, including its supported features and descriptions regarding the additions and modifications with respect to the original ns-3 modules from which it was generated, i.e., the \gls{lte}~\cite{BaldoLena} and the mmWave~\cite{nyummwave} ns-3 simulation models. We have designed the NR model by following as much as possible the latest 3GPP specifications, which we reference through the document when appropriate. Besides, we present some examples of usage and simulation results. We start in Section~\ref{sec:contrib} by giving a brief overview of the related work in the 5G simulation and emulation domain, as well as positioning our contribution with respect to the scientific community. Section~\ref{sec:overview} gives an overview of the NR simulator components, before entering into the details of the PHY layer in Section~\ref{sec:phy} and the MAC layer in Section~\ref{sec:mac}. In Section~\ref{sec:bwp}, our innovative work for managing different bandwidth parts is described. Then, we present the calibration procedure and an example of usage in a realistic 5G scenario in Section~\ref{sec:use}. Finally, Section~\ref{sec:future} discusses our roadmap and future plans, and Section~\ref{sec:conclusions} concludes the work.

\section{Scientific Contribution}
\label{sec:contrib}

In this section, we briefly analyze the existing available simulation tools, and then we highlight the novelty with respect to other works, and the contribution to the scientific community provided by our simulator.

\subsection{Related work}
A key challenge to perform new technology evaluations is that, despite the large body of results presented in the literature and produced by 3GPP evaluation working groups, the simulators are not publicly available. Usually, the obtained results are not reproducible, and system performance metrics are presented without much detail revealed about the underlying models and assumptions.  
Normally, simulators used by companies in 3GPP are required to pass through a calibration procedure, but they are private, and consequently not available to the research community. There are private commercial simulators that are available after paying an annual license fee for using them. Often, if not in all cases, the license is very restrictive and does not allow modifications or inspection of the source code, which is a clear limit for the research and the potential innovation. Our simulator, in turn, is GPLv2-licensed and guarantees the freedoms of the free software (free as in speech, not free as in beer) movement. We do not advocate for any political positions in this paper, but at the same time, we are interested in fostering the reproducibility of results, the collaborative development, and the support to the open innovation. Therefore, in the short review that follows, we only focus on software that is openly available and that guarantees, at least for academic purposes, the same freedom as we are guaranteeing.

The OpenAirInterface~\cite{nikaein2014openairinterface} is an open source platform for the simulation of wireless networks. Since April 2018, it supports the NR specifications. It can be used in a real testbed, but it misses a more comprehensive simulation setup that includes configuration and tracing of variables, as well as integration with other technologies to conduct interoperability experiments. Differently, our simulator is covered by the ns-3 umbrella, which includes models for multiple technologies like \gls{lte} and WiFi, among others, and therefore it offers the option of evaluating multi-\gls{rat} coexistence scenarios.

Concerning system-level simulators, an interesting software is the Vienna Simulator~\cite{vienna5g}. The research group from TU Wien developed a MATLAB tool that allows researchers to perform link- and system-level simulations for LTE and NR. In combination with several propagation models, the Vienna simulator allows simulating the network performance based on signal strength and accumulated interference. Generally speaking, the simulator is of interest to people working at the PHY and MAC layers. Differently, network simulators need to abstract the PHY layer through look-up tables to reduce the computational time and focus more on the MAC and higher layers.

The last category is the domain of the network simulation, in which our simulator is classified. Besides ns-3, from which we derive for the reasons that we explain in the next subsection, there is OMNeT++~\cite{virdis2015simulating}. OMNeT++ has support for LTE and LTE-Advanced features but lacks support for NR.

\subsection{Contribution}

The NR module starts as a fork of the ns-3 mmWave simulation tool developed by New York University (NYU) and University of Padova~\cite{nyummwave}.
The mmWave simulation model imports fundamental \gls{lte} features from the ns-3 LTE module (LENA)~\cite{BaldoLena}, which has been entirely designed and developed at Centre Tecnol\`ogic de Telecomunicacions Catalunya (CTTC). Our contributions entail a comprehensive and intensive work to align the mmWave module to the latest NR standard published by 3GPP. We have chosen as a base the popular ns-3 framework, as well as the mmWave simulation tool, for many reasons. 

On the one hand, ns-3 is an open-source discrete-event network simulator, and thus we inherit the capability of tracing internal events, a flexible configuration system, and a variety of modules to simulate other technologies, such as Ethernet, LTE, or WiFi, and multi-technology scenarios. As such, we can model different segments of the same network and, when needed, inter-connect different systems. The ns-3 simulator is widely adopted, recognized in research and academia, well maintained by an active community, and receives typically also support from the Google Summer of Code, as a flagship open source project. Thanks to ns-3, we also have the possibility of running our model in real time, therefore leaving simulation and entering the emulation domain with real equipment. We discussed the initial development of our simulator in~\cite{biljana:18}. 

On the other hand, we have chosen the mmWave simulation tool as our starting point because it includes already multiple features that are of interest for NR, mainly to access the mmWave spectrum above 6 GHz. In particular, the researchers from NYU and the University of Padova have done a solid job in modeling the aspects of beamforming, antenna gain, and propagation channel models, which makes the resulting work a milestone in the history of ns-3. However, the implementation of the mmWave module started in a moment in time when NR 3GPP specifications were not available, and the general vision of the technology was not as stable as it is today. As a result, many implemented aspects were not standard compliant and needed a revision, like the frame structure. Other things, such as the modeling of the channel and the beam management representation, were, on the contrary, entirely in line with the 3GPP standard, and therefore we have not modified them.

The mmWave module derives from the ns-3 LTE module (LENA)~\cite{BaldoLena}, so that both the mmWave and NR modules are also highly influenced by the previous design of the LTE module. In particular, both modules reuse from LTE all the higher protocol (\gls{rlc}, \gls{pdcp}, \gls{rrc}, \gls{nas}), as well as the \gls{epc}. However, we have done work to upstream the NR module to the ns-3 framework. As such, the NR module will further benefit from additions that future contributors will make to the ns-3 simulator, and it can reuse exciting features such as the Direct Code Execution~\cite{DCE}: users can perform simulations with realistic TCP/IP implementations and existing applications.
In addition, in the spirit of ns-3, and inherited from the LTE module, the NR model abstracts some low-level details, interpolating the values from a static set of look-up tables, to reduce computational aspects and facilitate simulating wider and complex scenarios with many base stations and users.

The rest of the paper is dedicated to explaining the design and implementation of our Non-Standalone (NSA) \gls{nr} simulator, which includes 4G \gls{epc} and 5G \gls{ran}. The main \gls{nr} features that we have added and modified to the mmWave tool are: 
\begin{itemize}
    \item flexible and automatic configuration of the \gls{nr} frame structure through multiple numerologies;
    \item \gls{ofdma}-based access with variable \gls{tti}s;
    \item restructuring and redesign of the \gls{mac} layer, including new flexible \gls{mac} schedulers that simultaneously consider time- and frequency-domain resources (resource blocks and symbols)  both for \gls{tdma} and \gls{ofdma}-based access schemes with variable \gls{tti};
    \item  \gls{ul} grant-based access scheme with scheduling request and 3GPP-compliant buffer status reporting;
    \item NR-compliant processing timings;
    \item new \gls{bwp} managers and the architecture to support operation through multiple \gls{bwp}s.
\end{itemize} 

\section{NR module overview}
\label{sec:overview}

\begin{figure}[!t]
  \centering
  \includegraphics[width=1\linewidth]{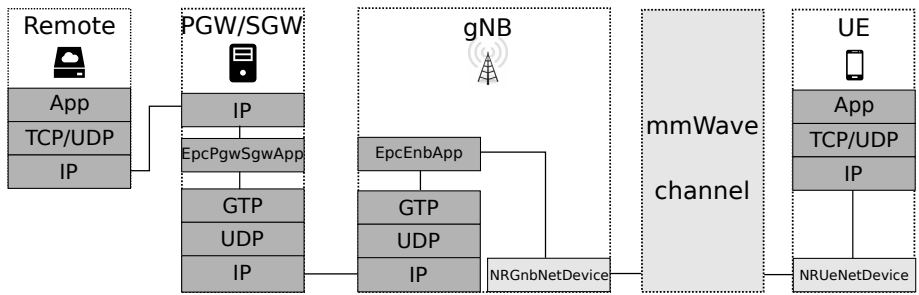}
  \caption{End-to-end class overview. In dark gray, we represent the existing, and unmodified, ns-3 and LENA components. In light gray, we represent the mmWave/NR components.}
  \label{fig:e2e-overview}
\end{figure}

We designed the \gls{nr} module to be able to perform \gls{e2e} simulations of 3GPP-oriented cellular networks. The \gls{e2e} overview of a typical simulation with the NR model is drawn in Figure~\ref{fig:e2e-overview}. On one side, we have a remote host (depicted as a single node in the figure, for simplicity, but there can be multiple nodes) that connects to an \gls{sgw}/\gls{pgw}, through a link. Such a connection can be of any technology that is currently available in ns-3. It is currently implemented through a single link, but it can be replaced by an entire subnetwork with many nodes and routing rules. Inside the \gls{sgw}/\gls{pgw}, the \cl{EpcSgwPgwApp} encapsulates the packet using the \gls{gtp} protocol. Through an IP connection, which represents the backhaul of the NR network (again, represented with a single link in the figure, but the topology can vary), the GTP packet is received by the \gls{gnb}. There, after decapsulating the payload, the packet is transmitted over the \gls{ran} through the entry point represented by the class \cl{NRGnbNetDevice}. The packet, if received correctly at the \gls{ue}, is passed to higher layers by the class \cl{NRUeNetDevice}. The path crossed by packets in the \gls{ul} case is the same as the one described above but on the contrary direction. We will detail our modifications to support NR in the PHY classes in Section~\ref{sec:phy}.

\begin{figure}[!t]
  \centering
  \includegraphics[width=0.50\linewidth]{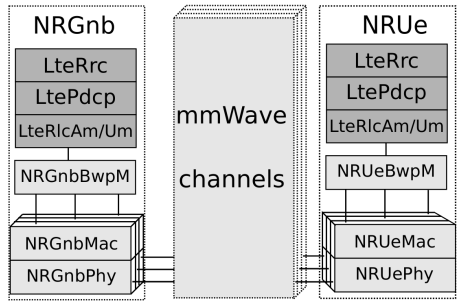}
  \caption{RAN class overview. In dark gray, we represent the existing, and unmodified, LENA components. In light gray, we represent the mmWave/NR components.}
  \label{fig:ran-overview}
\end{figure}

Concerning the \gls{ran}, we detail what is happening between \cl{NRGnbNetDevice} and \cl{NRUeNetDevice} in Figure~\ref{fig:ran-overview}. The \cl{NRGnbMac} and \cl{NRUeMac} \gls{mac} classes implement the LTE module \gls{sap} provider and user interfaces, enabling the communication with the LTE \gls{rlc} layer. The module supports \gls{rlc} \gls{tm}, \gls{sm}, \gls{um}, and \gls{am} modes. The \gls{mac} layer contains the scheduler (\cl{NRMacScheduler} and derived classes). Every scheduler also implements a \gls{sap} for LTE \gls{rrc} layer configuration (\cl{LteEnbRrc}). The \cl{NRPhy} classes are used to perform the directional communication for both \gls{dl} and \gls{ul}, to transmit/receive the data and control channels. Each \cl{NRPhy} class writes into an instance of \cl{MmWaveSpectrumPhy} class, which is shared between the \gls{ul} and \gls{dl} parts. We did not modify the internal of \cl{MmWaveSpectrumPhy} and, as for the original design of mmWave, it contains many PHY-layer models: interference calculation, \gls{sinr} calculation, the Mutual Information (MI)-based error model (to compute the packet error probability), as well as the Hybrid ARQ PHY-layer entity to perform soft combining. We will detail our modifications to support NR in the MAC classes in Section~\ref{sec:mac}.

Interesting blocks in Figure~\ref{fig:ran-overview} are the \cl{NRGnbBwpM} and \cl{NRUeBwpM} layers. \gls{3gpp} does not explicitly define them, and as such, they are virtual layers, but they help construct a fundamental feature of our simulator: the multiplexing of different \gls{bwp}s. NR has included the definition of \gls{bwp} for energy-saving purposes, as well as to multiplex a variety of services with different \gls{qos} requirements \cite[Sect. 6.10]{TS38300}. In our simulator, it is possible to divide the entire bandwidth into different \gls{bwp}s. Each \gls{bwp} can have its own \gls{phy} and \gls{mac} configuration (e.g., a specific numerology, scheduler rationale, and so on). We added the possibility for any node to transmit and receive flows in different \gls{bwp}s, by either assigning each bearer to a specific \gls{bwp} or distributing the data flow among different \gls{bwp}s, according to the rules of the \gls{bwp} manager. The introduction of a proxy layer to multiplex and demultiplex the data was necessary to glue everything together, and this is the purpose of these two new classes (\cl{NRGnbBwpM} and \cl{NRUeBwpM}). Everything briefly explained here will be analyzed in Section~\ref{sec:bwp}.

\section{Physical Layer}
\label{sec:phy}
One of the fundamental features that the NR simulator must support is the flexible frame structure defined in NR specifications~\cite{TS38300}.

LTE considers three types of frame structures: Type 1 for \gls{fdd}, Type 2 for \gls{tdd}, and Type 3 for use in the unlicensed spectrum (Licensed-Assisted Access). In general, each of these frames contains ten subframes, of 1 ms each. The ns-3 LTE model supports \gls{fdd} and focuses on Type 1 frame structure. In that model, the subframe is, as per~\cite{TS36211}, organized into two slots of 0.5 ms length. The model considers the implementation of the normal \gls{cp}, which results in seven \gls{ofdm} symbols per slot. The MAC scheduler of the ns-3 LTE model, however, only allocates resources within the subframe granularity in the \gls{pdsch} and \gls{pusch}.

Differently, NR introduces the concept of numerology, defined by a \gls{scs} and a \gls{cp}~\cite{NR-study-38.912-Rel14,TS38300}, and defines a set of numerologies to be supported. Currently, six frame structures are supported in the standard: 1) $\mu=0$ (\gls{scs}=15 KHz) and normal \gls{cp}, 2) $\mu=1$ (\gls{scs}=30 KHz) and normal \gls{cp}, 3) $\mu=2$ (\gls{scs}=60 KHz) and normal \gls{cp}, 4) $\mu=2$ (\gls{scs}=60 KHz) and extended \gls{cp}, 5) $\mu=3$ (\gls{scs}=120 KHz) and normal \gls{cp}, and 6) $\mu=4$ (\gls{scs}=240 KHz) and normal \gls{cp}~\cite[Sect. 4.3.2]{TS38211}.

The NR frame length is 10 ms, and an NR frame is composed of ten subframes of 1 ms each, to maintain backward compatibility with \gls{lte}. 
However, differently from \gls{lte}, each subframe is split in the time domain into a variable number of slots that depends on the configured numerology.
In particular, as the \gls{scs} increases, the number of slots per subframe increases and the slot length reduces. The number of \gls{ofdm} symbols per slot depends on the \gls{cp} length: 14 \gls{ofdm} symbols per slot for normal \gls{cp}, and 12 \gls{ofdm} symbols per slot for extended \gls{cp}. In the frequency domain, the number of subcarriers per \gls{prb} is fixed to 12 (as in \gls{lte}). Actually, for $\mu=0$, the NR frame structure is equal to that of \gls{lte}, although the concept of \emph{slot} has changed. Thus, according to NR specifications and differently from \gls{lte}, the NR frame structure is flexible and allows different \gls{ofdm} symbol lengths, slot lengths, and the number of slots per subframe. In addition, more flexibility is added to the NR MAC scheduler, which works on a numerology-dependent slot-basis (instead of a fixed 1 ms subframe-basis, as in LTE).

The ns-3 NR model implements TDD and supports the different NR numerologies, as shown in Table~\ref{table:tab-numerologies} for normal \gls{cp}. As it can be observed, differently from \gls{lte}, the \gls{scs}, the slot length, the \gls{ofdm} symbol length, and the CP length have different values depending on the numerology that is configured. Also, as the SCS varies with the numerology, the number of PRBs within the system bandwidth is numerology-dependent, as illustrated in Table~\ref{table:tab-numerologies}.

\begin{table*}[t!]
\centering
\footnotesize
\begin{tabular}{ |c|c|c|c|c|c|c| } 
 \hline
  & $\mu=0$ & $\mu=1$ & $\mu=2$ & $\mu=3$ & $\mu=4$\\
  \hline
 SCS [kHz] & 15 & 30 & 60 & 120 & 240\\
 \hline
 \gls{ofdm} symbol length [us] & 66.67 & 33.33 & 16.67 & 8.33 &  4.17\\
 \hline
 Cyclyc prefix [us] & $\sim$4.8 & $\sim$2.4 & $\sim$1.2 & $\sim$0.6 & $\sim$0.3\\
 \hline
 Number of subframes in frame & 10 & 10 & 10 & 10 & 10\\
 \hline
 Number of slots in subframe & 1 & 2 & 4 & 8 & 16\\
 \hline
 Slot length [us] & 1000 & 500 & 250 & 125 & 62.5\\
 \hline 
 Number of \gls{ofdm} symbols in slot & 14 & 14 & 14 & 14 & 14\\
 \hline
 Number of subcarriers in a PRB & 12 & 12& 12& 12& 12\\
 \hline 
 PRB width [MHz] & 0.18 & 0.36 & 0.72 & 1.44 & 2.88\\
 \hline
\end{tabular}
\caption{NR numerologies}
\label{table:tab-numerologies}
\end{table*}

To configure the numerologies in the NR module in a user-friendly way, we have a single numerology attribute value to be set in the class \cl{NrPhyMacCommon}, which refers to $\mu$. Then, differently to the original code, we derive the PHY layer parameters as follows:

\begin{itemize} \setlength\itemsep{0em}
  \item The number of slots per subframe ($n$) is: $n = 2^{\mu}$;
  \item The slot period ($t_s$) is set to: $t_s = \frac{1}{n}$ ms;
  \item The number of OFDM symbols per slot is fixed to 14, according to normal CP;
  \item The OFDM symbol period ($t_{os}$) is set to: $t_{os} = \frac{t_s}{14}$ ms;
  \item The number of subcarriers per resource block is fixed to 12, as per 3GPP specifications;
  \item The subcarrier spacing ($SCS$) is: $SCS=  2^{\mu} * 15 * 1000$ Hz;
  \item For a total bandwidth of $BW$, the number of resource blocks ($N$) is then: $N=\lfloor \frac{BW}{SCS * 12} \rfloor$, being $\lfloor . \rfloor$ the floor function.
\end{itemize}

Our implementation currently supports the numerologies\footnote{In \gls{nr} Rel-15, not every numerology can be used for every physical channel and signal: $\mu{=}4$ is not supported for data channels, and $\mu{=}2$ is not supported for synchronization signals~\cite{TS38300}. Also, for data channels, only $\mu{=}0,1,2$ are supported in frequency range 1 (sub 6 GHz, 0.45 - 6 GHz) and $\mu{=}2,3$ in frequency range 2 (mmWave, 24.25 - 52.6 GHz).} shown in Table~\ref{table:tab-numerologies}. It also supports $\mu=5$, which might be used in future NR releases for operation at high carrier frequencies. $\mu=5$ is defined by SCS of 480~kHz, \gls{ofdm} symbol length of 2.08~$us$, CP of 0.15~$us$, slot length of 31.25~$us$, and contains 32 slots in a single subframe.

\begin{figure}[t!]
\centering
\includegraphics[width=0.9\linewidth,draft=false]{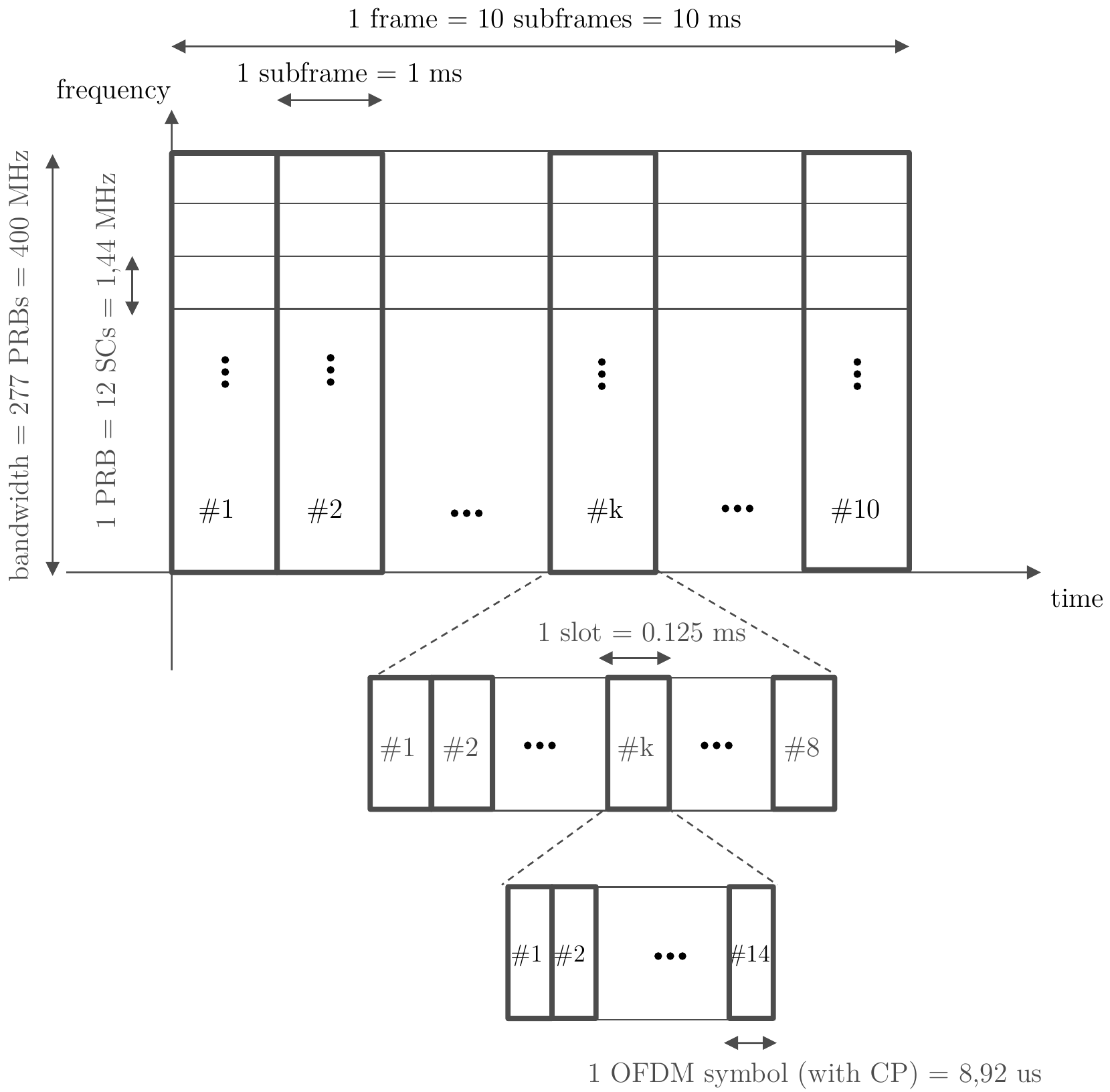}
\caption{NR frame structure in time- and frequency- domain for $\mu=3$ and a total channel bandwidth of 400 MHz.}
\label{fig:fig-frame-structure-time}
\end{figure}

In Figure \ref{fig:fig-frame-structure-time}, we illustrate the implemented NR frame structure in time- and frequency- domain when configured for $\mu=3$ (i.e., SCS=$120$~kHz) and a total channel bandwidth of 400~MHz. To simplify modeling in the simulator, the CP is included jointly with the \gls{ofdm} symbol length. For example, for $\mu=3$, the \gls{ofdm} symbol length including CP is $8.92$~$us$, which accounts for the real \gls{ofdm} symbol length 1/SCS=$8.33$~$us$ plus a CP of $0.59$~$us$.

To support a realistic NR simulation, we properly model (as per the standard) the numerology-dependent slot and \gls{ofdm} symbol granularity~\cite{biljana:18}. Differently from the original mmWave and LTE module (which only disposed of frame/subframe/symbol granularities) we have introduced the \emph{slot} granularity. Therefore, our code is able to support the NR frame structure in the time domain, and the scheduling operation per slot. Accordingly, we have adapted other parts of the simulator to the new NR frame structure: the \gls{phy} transmission/reception functionality, the \gls{mac} scheduling and the resource allocation information, the processing delays, and the interaction of the \gls{phy} layer with the \gls{mac} layer, as detailed in the following.

Firstly, the transmission and the reception have been updated to be performed on a slot basis. The corresponding functions are \cl{StartSlot}() and \cl{EndSlot}(), respectively. These functions are executed at both gNB and UE every 14 \gls{ofdm} symbols, so that the time periodicity depends on the configured numerology. According to the NR definition~\cite{NR-study-38.912-Rel14}, one TTI corresponds to many consecutive \gls{ofdm} symbols in the time domain in one transmission direction, and different TTI durations can be defined when using a different number of \gls{ofdm} symbols (e.g., corresponding to a mini-slot, one slot, or several slots in one transmit direction). Thus, the TTI is in general of variable length, regardless of the numerology. Transmission and reception of TTIs of variable length, as per NR, are handled by \cl{StartVarTti}() and \cl{EndVarTti}() functions, respectively. 

Secondly, to support operation per slot, we introduced \cl{SlotAllocInfo} that stores the resource allocation information at a slot level. A single \cl{SlotAllocInfo} includes a list of \cl{VarTtiAllocInfo} elements where each element contains the scheduling information per TTI, whose duration is no longer than that of one slot. For example, \cl{VarTtiAllocInfo} objects are populated at the UE after reception of \gls{dci} messages. For each \cl{VarTtiAllocInfo}, the \gls{mac} scheduler specifies the assigned PRBs and \gls{ofdm} symbols, along with the information whether the allocation is DL or UL, and whether it is control or data. More details regarding the scheduling process and the scheduling structures will be provided in Section~\ref{sec:mac}. 

Currently, as per the flexible slot structure in NR~\cite{TS38300}, the first and the last \gls{ofdm} symbols of the slot are reserved for DL control and UL control, respectively, while the \gls{ofdm} symbols in between can be dynamically allocated to DL or UL data. This enables dynamic TDD and also a flexible and configurable slot structure to allow fast DL-UL switch for bidirectional transmissions \cite[Sect. 4.3.2]{TS38211},  \cite{qualcomm:15b}.

Thirdly, regarding the processing delays, we have configured the \gls{mac}-to-\gls{phy} processing delay at the gNB to be numerology-dependent. Such delay is defined as a specific number of slots and configured by default to 2~slots. Differently, the transport block decoding time at the UE is by default fixed and equals to 100~$us$, but it could also be easily configured to be numerology-dependent. 

Finally, the slot inclusion at \gls{phy} requires also interaction with the \gls{mac} layer, because the \gls{mac} scheduling is performed per slot. Accordingly, a slot indication to the \gls{mac} layer has been included to trigger the scheduler at the beginning of each slot, to allocate a future slot. We enter into the details of the new NR schedulers in Section~\ref{sec:mac}.

\section{MAC Layer}
\label{sec:mac}

We have implemented the MAC layer in the classes \cl{NRGnbMac} and \cl{NRUeMac}. They interact directly with the \gls{phy} layer through a set of \gls{sap} APIs, and indirectly with the \gls{rlc} layer. The messages exchanged through the API between \gls{rlc} and \gls{mac} are captured and adequately routed by the bandwidth part manager (see more details about this in Section~\ref{sec:bwp}). As an example, the \gls{rlc} sends to the \gls{mac} many \gls{bsr} messages (one per bearer) to inform the scheduler of the quantity of data that is currently stored in the \gls{rlc} buffers. The scheduler, based on such information, takes then scheduling decisions. We have completely transformed the multiple access schemes, the \gls{ul} scheduling schemes, the scheduler timings, and the scheduler implementation part inside the \gls{mac} layer, as detailed in next subsections.

\subsection{Multiple Access Schemes}
We support \gls{ofdma} with variable \gls{tti} and \gls{tdma} with variable \gls{tti} schemes. In the case of \gls{ofdma}, we adapted the code to be able to assign a variable number of \gls{ofdm} symbols in time and \gls{rbg}s in frequency inside a slot. Visually, a \gls{tdma}-based scheme looks as depicted in Figure~\ref{fig:pure_tdma}. Three \gls{ue}s are scheduled, each one during a period of time that spans four \gls{ofdm} symbols and with data in all the \gls{rbg}s. A pure-\gls{ofdma} scheme allocates data of different \gls{ue}s on different \gls{rbg}s,  but using all the available \gls{ofdm} symbols, as shown in Figure~\ref{fig:pure_ofdma}. The new \gls{ofdma}-based scheme with variable \gls{tti}, instead, is the most flexible way to assign resources. It can allocate different \gls{rbg}s and a variable number of \gls{ofdm} symbols. 
An example is reported in Figure~\ref{fig:ofdma_var_tti}: \gls{ue}1 is allocated in a \gls{tdma} fashion in the first part of the slot, while \gls{ue}2 and \gls{ue}3 are scheduled in the rest of the \gls{ofdm} symbols, each one with a different set of \gls{rbg}s. Notice that while \gls{tdma} was available in the ns-3 mmWave module, and the pure-\gls{ofdma} was the original access supported in the ns-3 \gls{lte} module, we have extended the \gls{mac} layer to support both these schemes, plus the new \gls{ofdma}-based scheme with variable \gls{tti}.

In the \gls{nr} simulator, these multiple access schemes, as well as the scheduler policies for them, can be freely chosen (as we will explain later). However, it is worth noting that there are physical limitations when applying them to different spectrum regions. For instance, in the higher spectrum region (e.g., \gls{mmwave} bands) it would be more difficult to use the pure-\gls{ofdma} scheme due to incompatibility with the radio-frequency architectures that are based on single-beam capability~\cite{andrews:17}.

\def\RBGNum{20}
\def\SymNum{14}

\begin{figure*}[!t]
    \centering
    \begin{subfigure}[t]{0.30\textwidth}
\begin{tikzpicture}[scale=1, x=\linewidth/(\SymNum+3), y=10em / \RBGNum]
     
     \draw[fill=gray!15] (1,1) rectangle (2,20);
     \draw[fill=gray!15] (14,1) rectangle (15,20);
     \draw[preaction={fill=green!15}, pattern=north east lines, pattern color=black] (2,1) rectangle (6, 20);
     \draw[preaction={fill=yellow!15}, pattern=dots, pattern color=black] (6,1) rectangle (10,20);
     \draw[preaction={fill=blue!15}, pattern=north west lines, pattern color=black] (10,1) rectangle (14,20);
     
     \draw[black!100, step=1] (1,1) grid (\SymNum + 1, \RBGNum);
     \draw[thick, densely dashed, ->] (1, 1) -- (1, \RBGNum + 1);
     \draw[thick, densely dashed, ->] (1, 1) -- (\SymNum + 2, 1);
     \node [rotate=90] at (0, \RBGNum / 2) {Frequency (RBG)};
     \node [] at (8 , -1) {Time (Symbol)};

     \packet{2}{-5}{fill=gray!15}
     \node [] at (6, -4.5) {CTRL};
     
     \begin{scope}[shift={(9,-5)}]%
       \draw[preaction={fill=green!15}, pattern=north east lines, pattern color=black] (0,0) rectangle (1,1);
     \end{scope}
     \node [] at (12, -4.5) {UE1};
     
     \begin{scope}[shift={(2, -7)}]%
       \draw[preaction={fill=yellow!15}, pattern=dots, pattern color=black] (0,0) rectangle (1,1);
     \end{scope}
     \node [] at (6, -6.5) {UE2};
     
     \begin{scope}[shift={(9,-7)}]%
       \draw[preaction={fill=blue!15}, pattern=north west lines, pattern color=black] (0,0) rectangle (1,1);
     \end{scope}
     \node [] at (12, -6.5) {UE3};
\end{tikzpicture}

    \caption{Pure TDMA scheme.}
    \label{fig:pure_tdma}
    \end{subfigure}
    \hfill
    \begin{subfigure}[t]{0.30\textwidth}
    \begin{tikzpicture}[scale=1, x=\linewidth/(\SymNum+3), y=10em / \RBGNum]

     \draw[fill=gray!15] (1,1) rectangle (2,20);
     \draw[fill=gray!15] (14,1) rectangle (15,20);
  
     \draw[preaction={fill=green!15}, pattern=north east lines, pattern color=black] (2, 1) rectangle (14 , 7);
     \draw[preaction={fill=yellow!15}, pattern=dots, pattern color=black] (2, 7) rectangle (14 , 14);
     \draw[preaction={fill=blue!15}, pattern=north west lines, pattern color=black] (2, 14) rectangle (14, 20);
     
     \draw[black!100, step=1] (1,1) grid (\SymNum + 1, \RBGNum);
     \draw[thick, densely dashed, ->] (1, 1) -- (1, \RBGNum + 1);
     \draw[thick, densely dashed, ->] (1, 1) -- (\SymNum + 2, 1);
     \node [rotate=90] at (0, \RBGNum / 2) {Frequency (RBG)};
     \node [] at (8 , -1) {Time (Symbol)};

     \packet{2}{-5}{fill=gray!15}
     \node [] at (6, -4.5) {CTRL};
     
     \begin{scope}[shift={(9,-5)}]%
       \draw[preaction={fill=green!15}, pattern=north east lines, pattern color=black] (0,0) rectangle (1,1);
     \end{scope}
     \node [] at (12, -4.5) {UE1};
     
     \begin{scope}[shift={(2, -7)}]%
       \draw[preaction={fill=yellow!15}, pattern=dots, pattern color=black] (0,0) rectangle (1,1);
     \end{scope}
     \node [] at (6, -6.5) {UE2};
     
     \begin{scope}[shift={(9,-7)}]%
       \draw[preaction={fill=blue!15}, pattern=north west lines, pattern color=black] (0,0) rectangle (1,1);
     \end{scope}
     \node [] at (12, -6.5) {UE3};
\end{tikzpicture}

    \caption{Pure OFDMA scheme.}
    \label{fig:pure_ofdma}
    \end{subfigure}
    \hfill
    \begin{subfigure}[t]{0.30\textwidth}
    \begin{tikzpicture}[scale=1, x=\linewidth/(\SymNum+3), y=10em / \RBGNum]
     \draw[fill=gray!15] (1,1) rectangle (2,20);
     \draw[fill=gray!15] (14,1) rectangle (15,20);
     \draw[preaction={fill=green!15}, pattern=north east lines, pattern color=black] (2,1) rectangle (6, 20);
     \draw[preaction={fill=yellow!15}, pattern=dots, pattern color=black] (6,1) rectangle (14, 10);
     \draw[preaction={fill=blue!15}, pattern=north west lines, pattern color=black] (6, 10) rectangle (14, 20);
     
     \draw[black!100, step=1] (1,1) grid (\SymNum + 1, \RBGNum);
     \draw[thick, densely dashed, ->] (1, 1) -- (1, \RBGNum + 1);
     \draw[thick, densely dashed, ->] (1, 1) -- (\SymNum + 2, 1);
     \node [rotate=90] at (0, \RBGNum / 2) {Frequency (RBG)};
     \node [] at (8 , -1) {Time (Symbol)};

     \packet{2}{-5}{fill=gray!15}
     \node [] at (6, -4.5) {CTRL};
     
     \begin{scope}[shift={(9,-5)}]%
       \draw[preaction={fill=green!15}, pattern=north east lines, pattern color=black] (0,0) rectangle (1,1);
     \end{scope}
     \node [] at (12, -4.5) {UE1};
     
     \begin{scope}[shift={(2, -7)}]%
       \draw[preaction={fill=yellow!15}, pattern=dots, pattern color=black] (0,0) rectangle (1,1);
     \end{scope}
     \node [] at (6, -6.5) {UE2};
     
     \begin{scope}[shift={(9,-7)}]%
       \draw[preaction={fill=blue!15}, pattern=north west lines, pattern color=black] (0,0) rectangle (1,1);
     \end{scope}
     \node [] at (12, -6.5) {UE3};
\end{tikzpicture}

    \caption{OFDMA with variable TTI scheme.}
    \label{fig:ofdma_var_tti}
    \end{subfigure}
    
    \caption{Possible allocation schemes for a slot.}
    \label{fig:mac_allocations}
\end{figure*}
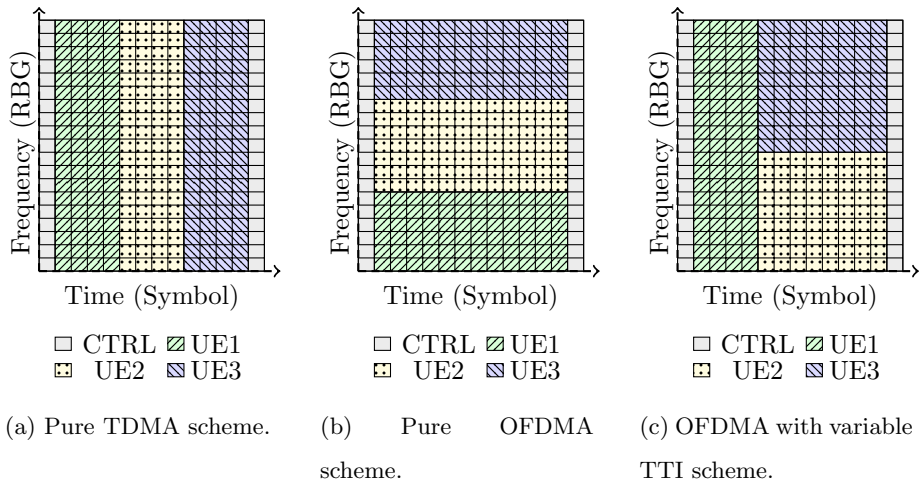

\subsection{Scheduling Schemes}
\gls{nr}, like \gls{lte}, uses dynamic scheduled-based access for \gls{dl}, based on which the gNB makes the scheduling decisions. Each \gls{ue} monitors the \gls{pdcch}, and upon the detection of a valid \gls{dci}, follows the given scheduling decision and receives its \gls{dl} data. In the case of \gls{ul}, \gls{nr} considers \gls{ul} grant-based and \gls{ul} grant-free access (also known as autonomous \gls{ul}) schemes~\cite{TS38300}. The former is the conventional dynamic scheduled-based access, as per LTE \gls{dl}/\gls{ul} and \gls{nr} DL, based on which the gNB makes the scheduling decisions in both UL and DL. Each \gls{ue} monitors the \gls{pdcch} and, upon the detection of a valid \gls{dci}, follows the given scheduling decision and transmits its \gls{ul} data. The latter is a contention-based scheme. At the time of writing, we have implemented only the \gls{ul} grant-based access, as per \gls{nr} specifications, but the \gls{ul} grant-free implementation is in our future roadmap. As a result, the \gls{nr} module supports dynamic scheduled-based accesses both for \gls{dl} and \gls{ul}.

The design that we followed aims to adopt different scheduling policies (round-robin, proportional fair, etc.) to a \gls{tdma} with variable \gls{tti}, a pure-\gls{ofdma}, or an \gls{ofdma} with variable \gls{tti} multiple access schemes. Also, we aim to reduce to the minimum the amount of duplicated code, while respecting the FemtoForum specification for LTE MAC Scheduler Interface. To do so, we considered that the primary output of a scheduler is a list of \gls{dci}s for a specific slot, each of which specifies (among other values) three crucial parameters. The first is the starting symbol, the second is the duration (number of \gls{ofdm} symbols), and the last one is the \gls{rbg} bitmap, in which a value of 1 in the position $m$ represents a transmission in the \gls{rbg} number $m$. This is compliant with \gls{dl} and \gls{ul} resource allocation Type 0 in \gls{nr}~\cite[Sect. 5.1.2.2 and Sect. 6.1.2.2]{TS38214}, as far as frequency-domain is concerned, and follows the standard time-domain resource allocation that includes \gls{sliv} for both \gls{dl} and \gls{ul}~\cite[Sect. 5.1.2.1 and Sect. 6.1.2.1]{TS38214}.

\begin{figure}[!t]
	\centering
	\includegraphics[width=0.98\linewidth]{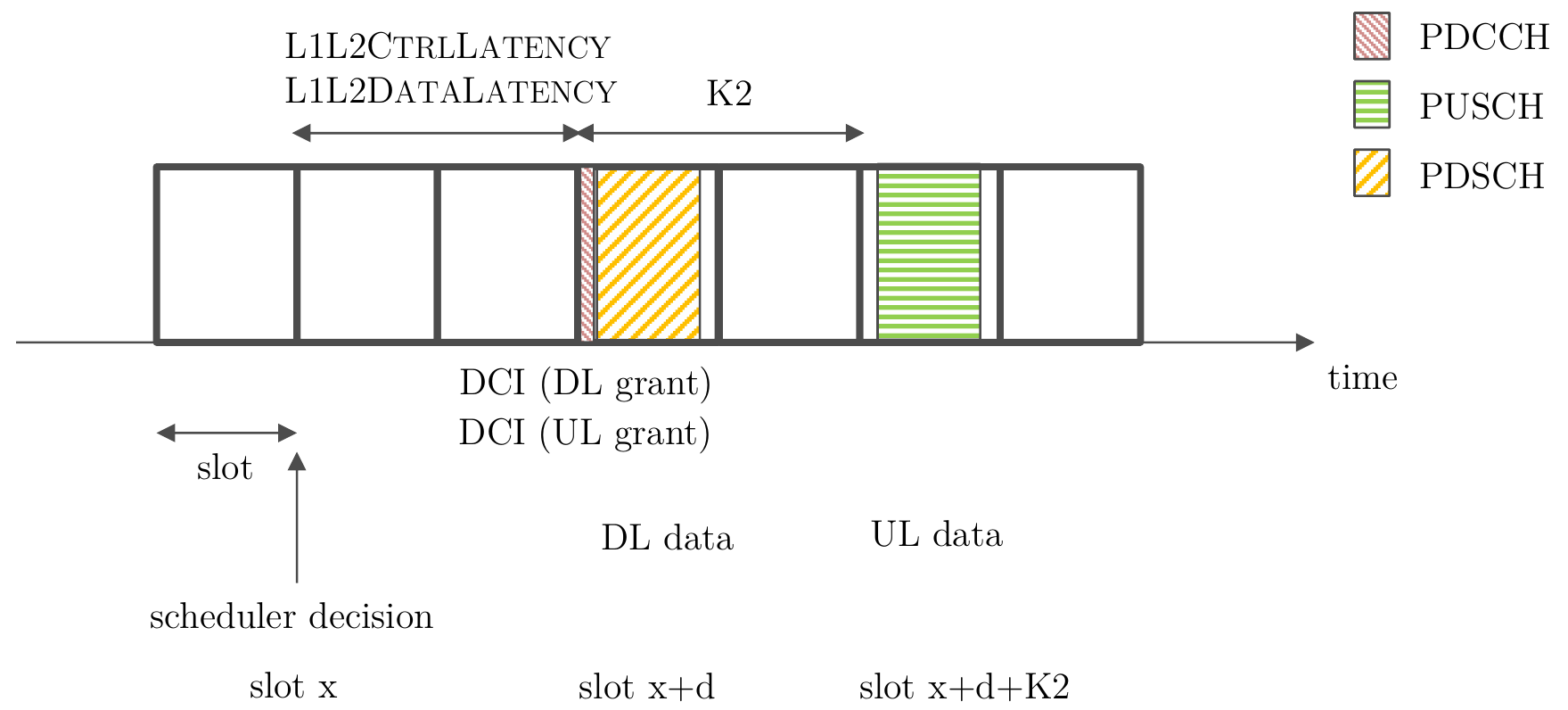}
	\caption{Scheduler timings in the ns-3 \gls{nr} simulator, for \kdos=2 slots and \attr{L1L2DataLatency}=\attr{L1L2CtrlLatency}=2 slots.}
	\label{fig_sched}
\end{figure}

\textbf{Scheduler Timings}:
We consider that the scheduler works ''ahead`` of time: at time $t$, when the PHY is transmitting slot $x$ over the air, the MAC is working to allocate slot $x + d$, where $d$ is a configurable delay, defined as a function of the number of slots. It represents the operational latency and, in the simulator, it is configured through the attribute \attr{L1L2CtrlLatency} and \attr{L1L2DataLatency} of the class \cl{NrPhyMacCommon}). For the \gls{dl} DCIs, this is the only delay to consider: when the slot $x + d$ is over the air, the \gls{dl} DCIs are transmitted in the first symbol and will apply for the same slot. However, for the \gls{ul} case, we must consider an additional delay which represents the time needed by the \gls{ue} to decode the \gls{dci} and to prepare the \gls{ul} data to transmit. The standard refers to this further delay as \kdos~\cite[Sect. 6.1.2.1]{TS38214}, which is measured in number of slots and can take any integer value from 0 to 32 slots. We model it through the attribute \attr{UlSchedDelay} of the class \cl{NrPhyMacCommon}. To keep it in consideration, if the \gls{phy} is transmitting over the air the slot $x$, the \gls{mac} will work on the \gls{ul} part of the slot $x + d +$ \kdos. These \gls{dci}s containing the \gls{ul} grant are transmitted over the air in slot $x + d$, and the \gls{ue} has \kdos~slots of time for preparing its \gls{ul} data. Figure~\ref{fig_sched} illustrates the scheduler operation and the DL/UL transmissions by taking into account these timings, for \kdos=2 slots and \attr{L1L2DataLatency} =\attr{L1L2CtrlLatency}=2 slots.

\textbf{UL handshake}: We have improved the dynamic scheduled-based access for \gls{ul} (i.e., the \gls{ul} grant-based scheme) as follows. Upon data arrival at the \gls{ue} \gls{rlc} queues, the \gls{ue} sends an \gls{sr} to the \gls{gnb} through the \gls{pucch} to request an \gls{ul} grant from its \gls{gnb}. Then, the \gls{gnb} sends the \gls{ul} grant (\gls{dci} in \gls{pdcch}) to indicate the scheduling opportunity for the \gls{ue} to transmit. Note that the first scheduling assignment is blind since the \gls{gnb} does not know the buffer size at the \gls{ue} yet. In this regard, since this is implementation-specific, we assume that the first scheduling opportunity consists of the minimum amount of \gls{ofdm} symbols that permits at least a 4 bytes transmission. In the majority of cases, this value equals to 1 \gls{ofdm} symbol. Next, the \gls{ue}, after receiving the \gls{ul} grant, performs the data transmission in the \gls{pusch}, which may contain \gls{ul} data and/or \gls{bsr}. After that, if a \gls{bsr} is received, the \gls{gnb} knows the \gls{ue} \gls{rlc} buffer status and can proceed with another \gls{ul} grant to account for the remaining data. Note that the main difference in the \gls{nr} module with respect to \gls{mmwave} and \gls{lte} ns-3 modules is that we have introduced the \gls{sr} in the \gls{pucch} and the \gls{bsr} can only be sent in conjunction with the \gls{mac} \gls{pdu} (since, according to \gls{nr} specifications, the \gls{bsr} is part of the \gls{mac} header), while in previous ns-3 modules the \gls{bsr} was sent periodically and ideally.

\begin{figure*}[!t]
	\centering
	\includegraphics[width=0.98\linewidth]{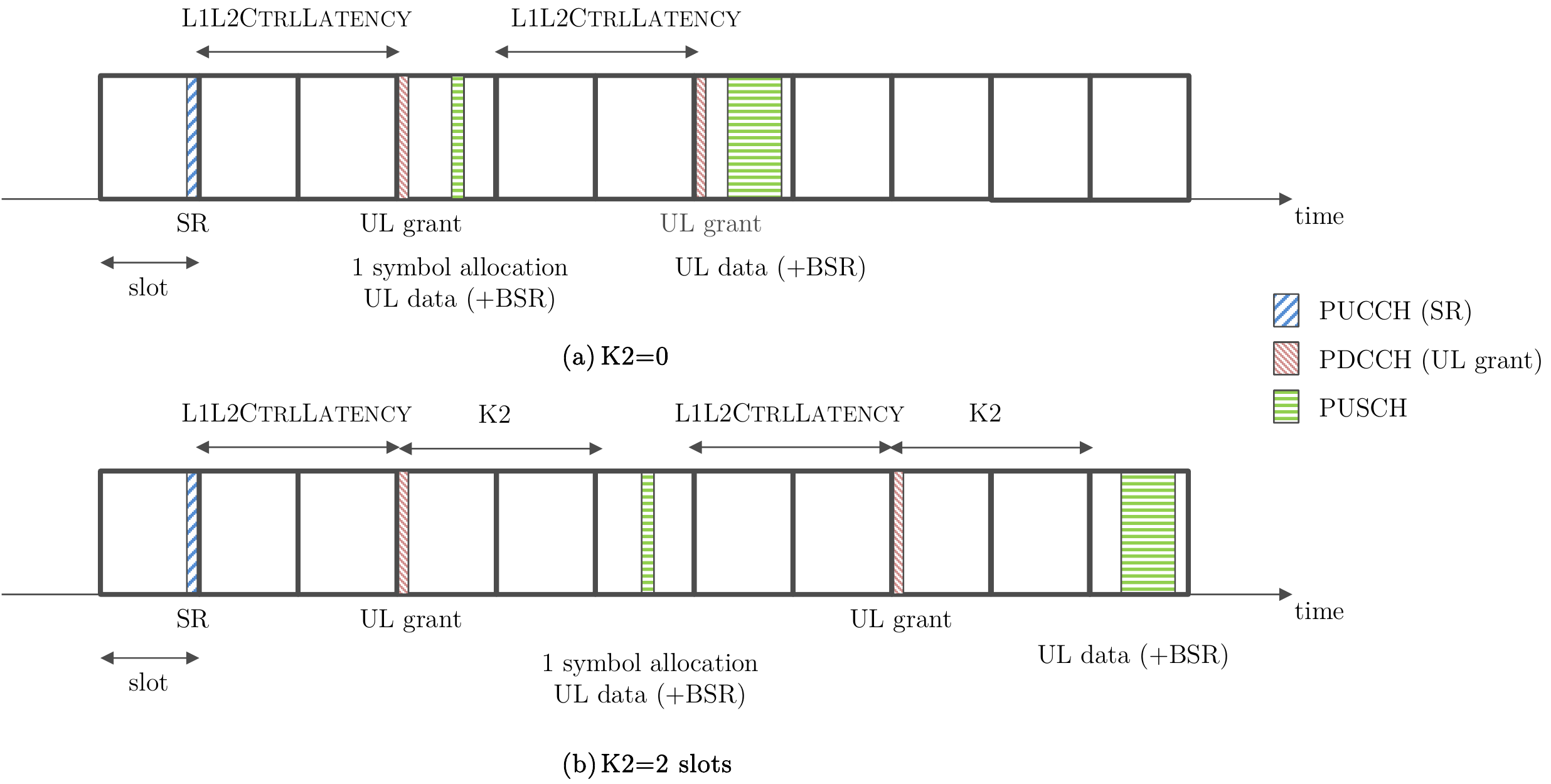}
	\caption{UL handshake procedure for \gls{nr} \gls{ul} grant-based access, including \gls{nr} timings and processing delays, as implemented in the ns-3 \gls{nr} simulator. (a) \kdos=0, (b) \kdos=2 slots.}
	\label{fig_UL}
\end{figure*}

Before sending the \gls{ul} grants, \attr{L1L2CtrlLatency} delay has to be considered at the \gls{gnb} side. Also, upon reception of an \gls{ul} grant, the \gls{ue} should send \gls{ul} data and/or \gls{bsr} after \kdos~slots, being \kdos~indicated in the \gls{ul} grant. So, these two parameters influence the \gls{ul} handshake. In Figure~\ref{fig_UL}, we show the \gls{ul} handshake, including also the timings and processing delays that influence it (i.e., \attr{L1L2CtrlLatency} and \kdos) for \kdos=0 (top) and \kdos=2 slots (bottom). Recall that we have a \gls{tdd} slot structure with 14 \gls{ofdm} symbols per slot, in which \gls{pdcch} is sent in the 1st symbol, \gls{pucch} in the 14th \gls{ofdm} symbol, and the \gls{ofdm} symbols in between are devoted to shared channels that may contain data (\gls{pdsch} and/or \gls{pusch}). Our implementation in the ns-3 \gls{nr} simulator follows exactly the handshake and timings that are illustrated in Figure~\ref{fig_UL}. The \gls{bsr} is prepared shortly before the \gls{phy} transmission in the \gls{ul}, reflecting the status of the \gls{rlc} queue without including the current transmission. 

\subsection{Scheduler Policies and Implementation at the \gls{gnb}}
The core class of the \gls{nr} schedulers design is \cl{NrMacSchedulerNs3}. This class defines the core scheduling process and splits the scheduling logic into logical blocks. The FemtoForum API splits the \gls{ul} and the \gls{dl} scheduling. In the following, we will consider only the \gls{dl}, but the description also applies to the \gls{ul} case. The differences lie in the variable and function naming, as well as the delays involved, as explained before. As a starting point, we prepare a list of active \gls{ue} and their requirements, organized based on the concrete beams they belong to. 

We start with the scheduler implementation details for \gls{ofdma}-based schemes. The first step of the procedure consists of distributing \gls{ofdm} symbols among multiple beams. We need this block for the \gls{ofdma}-based schemes because we chose to support single-beam capability only. At high frequencies, the beam is shaped after digital-to-analog conversion due to limitations in the implementation phase. Therefore, with analog beamforming, there is the constraint that a receive or transmit beam can only be formed in a single direction at any given time instant, meaning that if we want to transmit towards two \gls{ue}s with different beams, we must do so in different time instants. We provide two different ways to assign symbols to the beams: in a load-based or round-robin fashion. We consider as the beam load the sum of the bytes queued in the \gls{rlc} layer of the \gls{ue}s that belong to that beam. The round-robin assignment merely assigns the same number of \gls{ofdm} symbols to all beams.

After the symbols/beam selection in \gls{ofdma} schedulers, it is necessary to distribute the available \gls{rbg}s in the time/frequency domain among active \gls{ue}s in each beam. This step depends on the specific scheduling algorithm that the user has chosen. The \gls{rbg}s can be distributed following a round-robin, proportional fair, or maximum rate algorithm. The resources to be allocated are groups of \gls{rbg}s spanned over one, or more, symbols.

Finally, the last step consists in the creation of the corresponding DCI, based on the number of assigned resources made in the previous block. The assigned \gls{rbg}s should be grouped to create a single block for each \gls{ue}. Then, the \gls{rbg} bitmap is created\footnote{The ns-3 NR module follows NR resource allocation Type 0, as per~\cite{TS38214}, in which the resource allocation is specified through a bitmap. It provides more flexibility to the scheduler operation, as compared to NR resource allocation Type 1 that specifies the PRB start and number of PRB allocated.}, so that DCIs for different \gls{ue}s do not overlap. The bitmap will be an input, later on, for the \gls{phy} layer. At the transmission or reception time, the \gls{phy} translates the bitmap into a vector of enabled \gls{prb}. As the standard indicates in~\cite[Sect. 5.1.2.2 and 6.1.2.2]{TS38214}, each \gls{rbg} is grouping 2, 4, 8, or 16 \gls{prb} depending on the \gls{bwp} size. Then, the transmitter distributes the power, and the receiver decodes, only among the active \gls{prb}s.

The design also takes into consideration HARQ retransmissions. They have a higher priority in the scheduling policies. When a NACK is received, the scheduler takes the old DCI and tries to put it in the current slot for retransmission. If that is not possible, then it will be queued for the next slot. It is important to remark that the simulator only supports a round-robin policy to select the HARQ process to retransmit.

The user can select different schedulers and different assignment modes by swapping class name during the configuration phase. The available OFDMA schedulers are \cl{NrMacSchedulerOfdmaRR} (round-robin), \cl{NrMacSchedulerOfdmaPF} (proportional fair), and \cl{NrMacSchedulerOfdmaMR} (maximum rate). Our OFDMA schedulers are all using the variable TTI strategy, so they are allowed to create TTIs of different length. The configuration into pure OFDMA schedulers is straightforward.

For \gls{tdma}-based schedulers, the first step (symbols/beam selection) is not performed, as entire symbols are assigned to the UEs and the \gls{phy} layer is then perfectly capable of switching the beam in time (under the single-beam capability assumption explained before). Therefore, the assignment phase, in which the scheduler decides how many \gls{ofdm} symbols are assigned to each UE, is directly executed. We support round-robin (\cl{NrMacSchedulerTdmaRR}), proportional fair (\cl{NrMacSchedulerTdmaPF}), and maximum rate (\cl{NrMacSchedulerTdmaMR}) schedulers.

These classes, no matter the access mode, follow the same principles:
\begin{itemize}
  \item \emph{Round-robin}: The scheduler evenly distributes the available \gls{rbg}s among \gls{ue}s associated with that beam (\gls{ofdma}), while for \gls{tdma} evenly distributes the available symbols.
  \item \emph{Proportional fair}: In the \gls{ofdma} mode, the scheduler evenly distributes the available \gls{rbg}s among \gls{ue}s according to a metric that considers the actual rate, based on the \gls{cqi}) elevated to $\alpha$ and the average rate that has been provided in the previous slots to the different \gls{ue}s. By changing the $\alpha$ parameter the metric also changes. For $\alpha = 0$, the scheduler selects the \gls{ue} with the lowest average rate. For $\alpha = 1$, the scheduler selects the \gls{ue} with the largest ratio between actual rate and average rate. For \gls{tdma}, the resources to distribute are entire symbols.
  \item \emph{Maximum rate}: The scheduler evenly distributes the available \gls{rbg}s (or the available symbols in case of \gls{tdma}) among \gls{ue}s according to a maximum rate metric that considers the actual rate (based on the \gls{cqi}) of the different \gls{ue}s.
\end{itemize}

In the \gls{ul}, we currently support only \gls{tdma}. This means that, even for \gls{ofdma} schedulers, such a phase is treated as it was in the \gls{tdma} schedulers.

\section{Bandwidth Part Manager}
\label{sec:bwp}
\begin{figure}
    \centering
\begin{tikzpicture}[x=0.8cm,y=0.8cm]

    \draw[preaction={fill=blue!25}, pattern=north east lines] (0, 0) rectangle (6, 3);
    \draw[preaction={fill=red!25}, pattern=north west lines]  (0, 3) rectangle (6, 5);

		\draw[black!75] (0, 0) grid[xstep=2,ystep=0.5] (6, 3);
		\draw[black!75] (0, 3) grid[xstep=1,ystep=1] (6, 5);

		\draw[thick, densely dashed, <->] (-1, 0) -- node [rotate=90, above] {Ch. Bandwidth (B)} (-1, 5);

    \draw[thick, <->] (-0.25, 0) -- node [rotate=90, above] {$\mu$=1} (-0.25 , 3);
		\draw[thick, <->] (-0.25, 3) -- node [rotate=90, above] {$\mu$=2} (-0.25 , 5);

		\draw[thick, <->] (0, 5.25) -- node [above] {slot $\mu$=2} (1 , 5.25);
		\draw[thick, <->] (0, -0.25) -- node [below] {slot $\mu$=1} (2 , -0.25);

		\node[preaction={fill=blue!25}, rectangle, pattern=north east lines, label=below:BWP$_{1}$, inner sep=0.25cm] at (2, -1) {};
		\node[preaction={fill=red!25}, rectangle, pattern=north west lines, label=below:BWP$_{2}$, inner sep=0.25cm] at (4, -1) {};

\end{tikzpicture}
    \caption{An example of BWPs configuration.}
    \label{fig:bwps}
\end{figure}
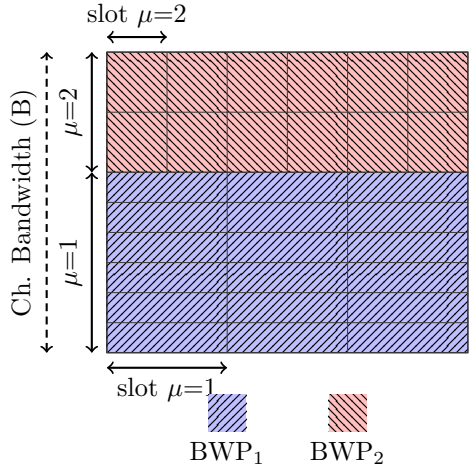

An additional level of flexibility in the NR system can be achieved by implementing the multiplexing of numerologies in the frequency domain. As an example, \gls{urllc} traffic requires a short slot length to meet strict latency requirements, while \gls{embb} use case in general aims at increasing throughput, which is achieved with a large slot length~\cite{zaidi:16}. Therefore, among the set of supported numerologies for a specific operational band and deployment configuration, \gls{urllc} can be served with the numerology that has the shortest slot length, and \gls{embb} with the numerology associated to the largest slot length~\cite{sandra_ref_3}. That is, the numerology for \gls{urllc} is recommended to be larger than the numerology for \gls{embb}, $\mu_u > \mu_e$, where $\mu_u$ is numerology used for \gls{urllc} and $\mu_e$ for \gls{embb}. Hence, the main objective of \gls{fdm} of numerologies is to address the trade-off between latency and throughput for different types of traffic by physically dividing the bandwidth in two or more \gls{bwp}s. In Figure \ref{fig:bwps}, we illustrate an example of \gls{fdm} of numerologies. The channel is split into two \gls{bwp} that accommodate the two numerologies ($\mu_u$ and $\mu_e$) multiplexed in frequency domain. The total bandwidth $B$ is then divided into two parts of bandwidth $B_u$ for \gls{urllc} and $B_e$ for \gls{embb}, so that $B_u + B_e \leq B$. The number of PRBs for \gls{urllc} is $N_u$ and $N_e$ for \gls{embb}. Note, that the PRB width varies with the numerology.


\subsection{FDM Model}
In the ns-3 NR simulator, the user can configure FDM bands statically before the simulation starts. This is a critical design assumption based on two main reasons. First, the NR module relies on the channel and the propagation loss model defined in the mmWave module, which is not able to allow runtime modifications of the physical configuration parameters related to time/frequency configuration (such as the system bandwidth, the central carrier frequency, and the symbol length). Thus, until the current mmWave/NR module channel model is not modified to allow these runtime configuration changes, it will not be possible to perform semi-static reconfiguration of BWPs. The second reason is that in the simulator the RRC messaging to configure the default bandwidth part, as well as the bandwidth part reconfiguration, are not implemented yet.


Regarding the data plane, there is a similarity among the concept of the component carriers (CCs) in LTE and the BWPs in NR. While the objective is different, since LTE aims to aggregate narrower bandwidth to achieve a wider capacity, while NR  \gls{bwp} intend to subdivide the bandwidth to use it for multiple and different purposes, the idea of having aggregated PHY layer instances remains the same. The main difference, at the implementation level, is that in LTE, the various carriers have the same \gls{ofdm} symbol, subframe, and frame boundary, while in NR only the subframe and frame boundaries of different BWPs are aligned, the slot and \gls{ofdm} symbol may not be. Everything else can be different, and therefore we can have contiguous \gls{bwp}s with different \gls{phy} parameters.

Following the previous discussion, it naturally comes that one possible way to implement the \gls{fdm} is to start from the \gls{ca} feature of ns-3 LTE~\cite{bojovicca}. This is exactly what we have done. In this line, the models of both \cl{NrGnbDevice} and \cl{NrUeDevice} have been extended to support the installation of instances of \gls{mac} and \gls{phy} per carriers, following the design (and inheriting the architecture) of the ns-3 LTE CA feature.

In the current implementation, we support the transmission of the scheduling information through a dedicated control channel in each \gls{bwp}, and the MAC scheduling and HARQ processes are performed per \gls{bwp}. Finally, according to our model, the multiplexing of the data flows based on the type of traffic is performed by a new layer, which is implemented by an entity called \cl{BwpManager}. Its role is similar to that of CC manager in the LTE module, and \cl{BwpManager} can use 5G \gls{qos} identifiers (as defined in~\cite{sandra_ref_8}) to determine on which \gls{bwp} to allocate the packets of a radio bearer and to establish priorities among radio bearers.


\subsection{Implementation of FDM of Numerologies}
\label{fdm_impl}
The main challenges to implement \gls{fdm} of numerologies lie in the modifications of the NR \gls{sap} interfaces between the layers of the NR stack to include the new \gls{bwp} layer according to the \gls{fdm} design. The basic block is a CC, which we will use as a synonym of \gls{bwp}. A CC consists of two instances, one for the \gls{mac} and the other for the \gls{phy} layer. Moreover, for each \gls{phy} there is a separate channel model, that is shared among the attached UEs. Each CC/BWP can be configured independently with different parameters (for instance, each \gls{phy} instance can be configured with a different transmit power) but the configuration should match with that of the attached UEs. In other words, the \gls{bwp}s configuration for the \gls{gnb} and the \gls{ue}s should be the same. Practically speaking, when the final user configures one or more \gls{bwp}s for one gNB, the code is automatically creating CCs and the configuration for the \gls{gnb} and all the attached UEs.

Then, we have a BWP Manager entity that is responsible for routing the traffic and the signaling messages to the correct BWP, based on their QoS requirements. Currently, this class is located above all the instances of \gls{mac} and \gls{phy} in UEs' and gNBs' stack. Each message that arrives at each CC is automatically redirected to the BWP Manager class, \cl{NRGnbBwpM}. The current implementation of \cl{NRGnbBwpM} supports all LTE EPS bearer QoS types, and the assignment of the corresponding BWP is based on the static configuration provided by the user that maps a bearer to a specific bandwidth part. So, the routing job is done through a lookup table, in which each message is mapped into an identifier, and then passed to the CC with that identifier. Similarly to the changes in gNB device, we have also extended the NR UE model to support the CCs and UE CC manager, to be able to route the \gls{ul} traffic properly. A visual representation of this class hierarchy is depicted in Figure~\ref{fig:ran-overview}, presented when we explained the overall architecture of the simulator.

The mmWave module channel models (MmWaveBeamforming, MmWaveChannelMatrix, MmWaveChannelRaytracing, MmWave3gppChannel) depend on the various PHY and MAC configuration parameters specified through a single instance of the class \cl{MmWavePhyMacCommon}. Hence, it is necessary to install as many \cl{MmWave3gppChannel} channel model instances as BWPs to configure. However, an important limitation of this design is that all gNBs and UEs in the simulation have the same BWP configuration. Also, the model does not consider interference between BWPs/CCs, so that appropriate band guards are to be left between contiguous BWPs/CCs.

Note that the architecture of the current implementation can be used either as a way to implement \gls{ran} slicing with a dedicated resource model~\cite{ksentini:17}, by allocating different flows to orthogonal \gls{bwp}s, or can be reused, with slight modifications, as a way to implement Carrier Aggregation~\cite{bojovicca}. Those are different use cases, with the same implementation blocks, for which the only difference lies in the logic of the BWP Managers at gNB and UE sides.

\section{Use Cases}
\label{sec:use}

In this section, we report two important simulation outcomes that we have obtained using the simulator. First, we discuss the validation of the models, following 3GPP calibration procedures. This ensures that the simulator, besides being properly tested with both system and unit tests, provides expected results as compared to those achieved by similar proprietary simulators. Second, we evaluate the E2E performance, as a function of different numerologies, in the context of a complex realistic 5G scenario, in order to show the potentiality of the simulator.

\subsection{Calibration}

The accuracy of the simulation results is very important when the simulator is used as a base to take design decisions or when it is needed to evaluate the effectiveness of a proposal. 3GPP holds calibration campaigns to align both link-level and system-level simulation results of different simulators. To align our simulator, we have followed the Indoor Hotspot (InH) system-level calibration for multi-antenna systems, as per~\cite[Annex A.2]{TR38802}. Details of the evaluation assumptions for Phase 1 NR MIMO system-level calibration are provided in~\cite{R1-1703534}, with further clarifications in~\cite{R1-1700144}, and are summarized in Table~\ref{table:sim_params}.

As reference curves, we use the results provided by the companies in~\cite{R1-1709828}. We consider the Cumulative Distribution Function (CDF) of the wideband SINR with beamforming, and the CDF of the wideband SNR with step b (i.e., with analog TX/RX beamforming, using a single digital TX/RX port). For each case, we depict as reference the average of the companies contributing to 3GPP, as well as the company that gets the minimum and the maximum of the average wideband SNR/SINR, so that an optimal region for calibration is defined. As a scenario, we are using the standard 3GPP calibration deployment, composed by 12 gNBs, deployed in two rows of six gNBs each, equally spaced by 20 meters. Horizontally, between every gNBs there are 20 m. Then, we locate 120 UEs (100$\%$ indoor) that are randomly dropped in a 50 m $\times$ 120 m area.

\begin{table}[t]
\footnotesize
\centering
\begin{tabular}{|c|c|}
\hline
Parameter       & Value                                                   \\ \hline
Carrier freq.   & 30 GHz                                                  \\ \hline
Bandwidth       & 40 MHz                                                  \\ \hline
SCS             & 60 kHz ($\mu$ = 2)                                      \\ \hline
Channel         & Indoor TR 38.900                                        \\ \hline
BS Tx Power     & 23 dBm                                                  \\ \hline
BS Antenna      & M=4, N=8, 1 sector, height=3 m, vertical polarization  \\ \hline
UE Antenna      & M=2, N=4, 1 panel, height=1.5 m, vertical polarization \\ \hline
BS noise figure & 7 dB                                                    \\ \hline
UE noise figure & 10 dB                                                   \\ \hline
UE speed        & 3 km/h                                                  \\ \hline
Scheduler       & TDMA PF                                                 \\ \hline
Traffic model   & Full Buffer                                             \\ \hline
\end{tabular}
\caption{Simulation parameters for calibration experiments}
\label{table:sim_params}
\end{table}

\begin{figure}[!t]
	\centering
	\includegraphics[width=0.8\columnwidth]{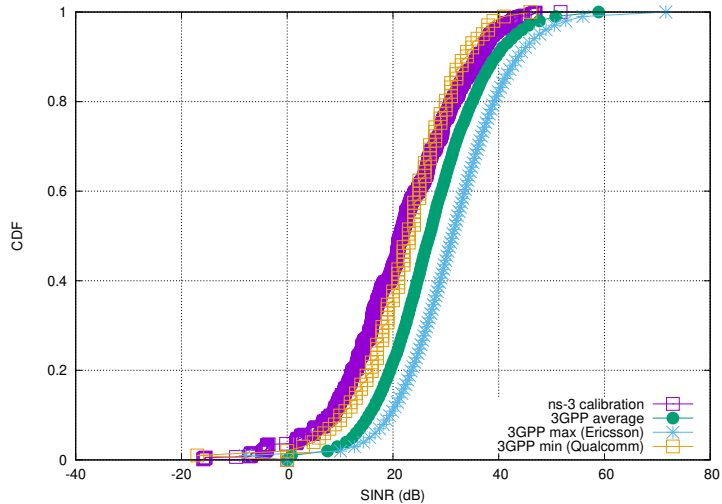}
	\caption{SINR for the case OfficeOpen with shadowing disabled.}
	\label{fig:OpenWall_SINR}
\end{figure}

\begin{figure}[!t]
	\centering
	\includegraphics[width=0.8\columnwidth]{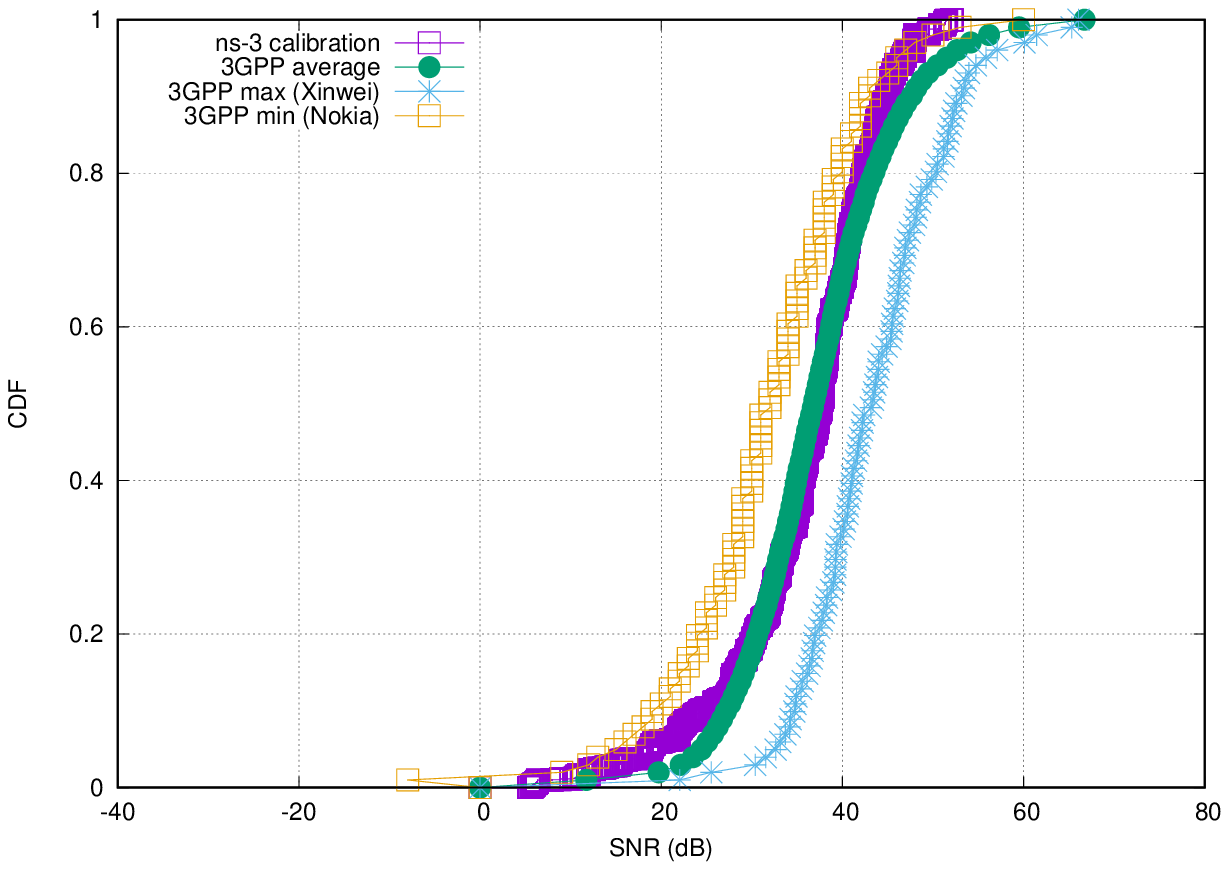}
	\caption{SNR for the case OfficeOpen with shadowing disabled.}
	\label{fig:OpenWall_SNR}
\end{figure}

\begin{figure}[!t]
	\centering
	\includegraphics[width=0.8\columnwidth]{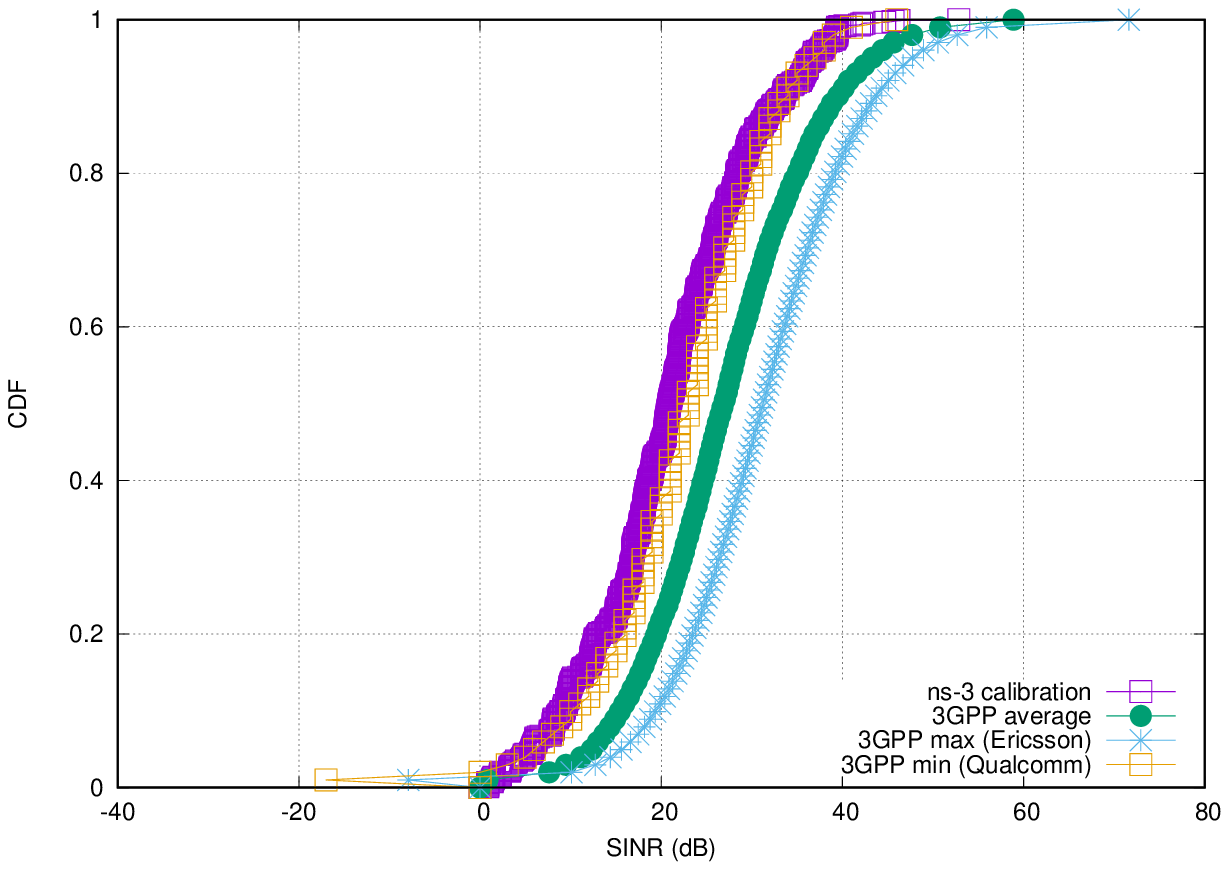}
	\caption{SINR for the case ShoppingMall with shadowing enabled.}
	\label{fig:ShoppingMall_SINR}
\end{figure}

\begin{figure}[!t]
	\centering
	\includegraphics[width=0.8\columnwidth]{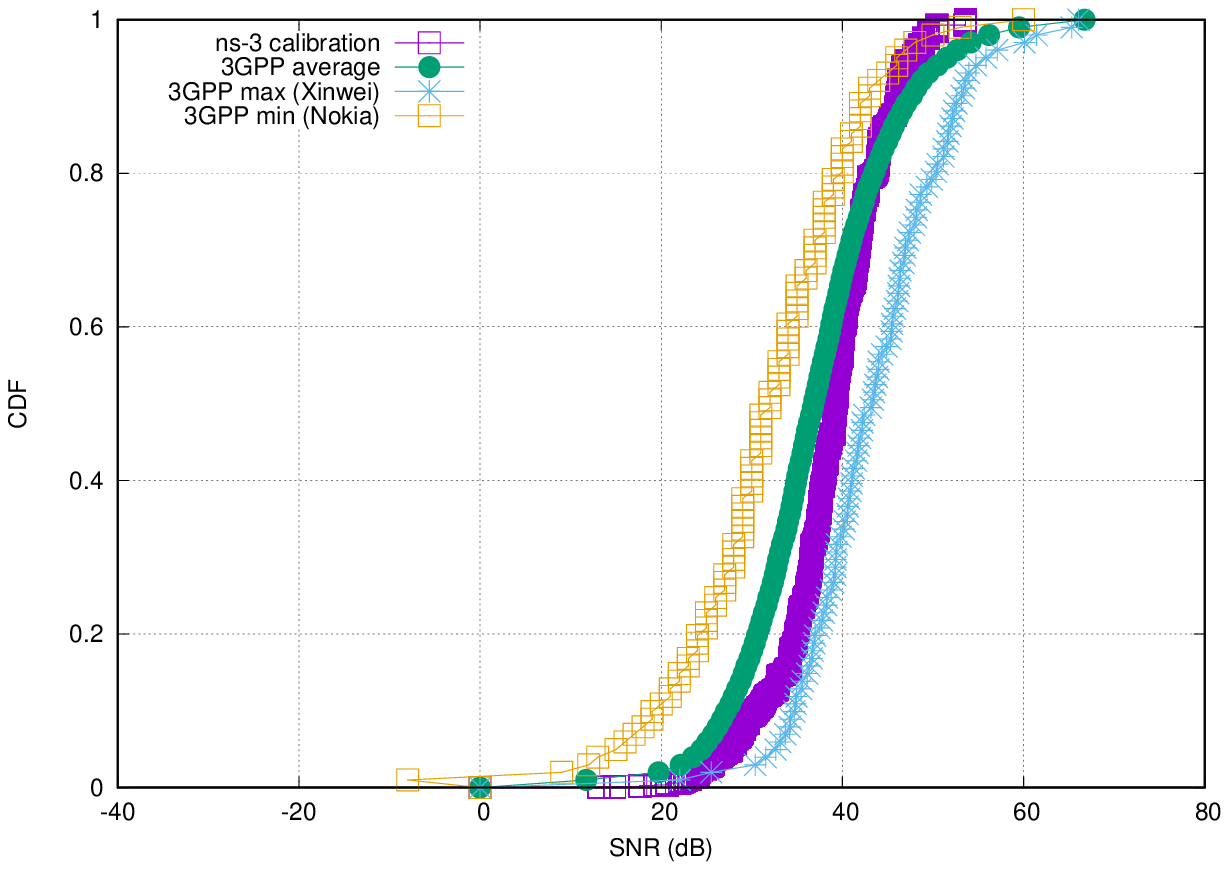}
	\caption{SNR for the case ShoppingMall with shadowing enabled.}
	\label{fig:ShoppingMall_SNR}
\end{figure}

In Figure~\ref{fig:OpenWall_SINR} and Figure~\ref{fig:OpenWall_SNR} we show the SINR and SNR, respectively, for the InH Office-Open propagation model, with shadowing disabled.
In Figure~\ref{fig:ShoppingMall_SINR} and Figure~\ref{fig:ShoppingMall_SNR}, we depict the SINR and SNR, respectively, of the InH Shopping-Mall propagation model, with shadowing enabled. In both cases, we can observe that the SINR is close to the lowest 3GPP reference curve. Regarding SNR, it lies entirely within the calibration region, with a perfect match with the average 3GPP SNR in the first configuration setup.

\subsection{E2E Latency and NR Numerologies}
In this subsection, we analyze a complex and realistic \gls{5g} future scenario, where traffic from multiple \gls{5g} applications is transmitted, using both UDP and TCP transport protocols. We study the impact of processing and decoding delays for differently configured numerologies and analyze how they affect the E2E performance.

\begin{figure}[!t]
  \centering
  \includegraphics[width=.95\columnwidth]{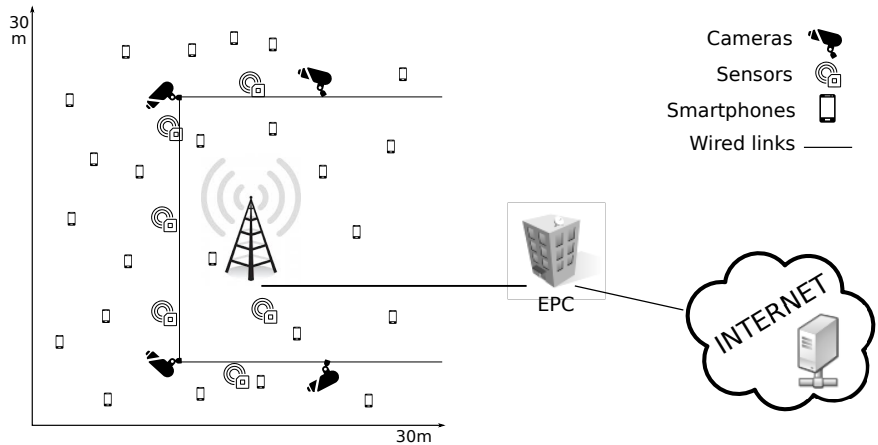}
  \caption{Reference scenario for analyzing E2E latency and NR numerologies.}
  \label{fig:reference_scenario}
\end{figure}

To model a real-world scenario, we base our simulation on the setup shown in Figure~\ref{fig:reference_scenario}. At a high level, we have a backbone connection between the \gls{epc} to remote nodes, modeled as 100 Gb/s point-to-point link. The link between the \gls{gnb} and the \gls{epc} that represents the \gls{cn} is made with another point-to-point connection with a maximum rate of 10 Gb/s, without propagation delay.

Regarding the \gls{ran}, we consider the use case of a next-generation school, served by a single \gls{gnb}, in which different but connected objects share the connectivity. We have twenty-five smartphones, six sensors, four IP cameras distributed over a circular area of 30 m of diameter. The position of each \gls{ue} in the reference scenario depicted in Figure~\ref{fig:reference_scenario} is indicative because in the simulations we have located the \gls{ue}s in random positions to gather more statistical significance in the results.

\begin{table}[!t]
\footnotesize
\centering
\begin{tabular}{|c|c|c|c|c|c|}
\hline
             & \# \textit{Flows} & \textit{Start} (s) & \textit{Rate (Mb/s)}  & \textit{Pkt Size (B)} & \textit{RAN Dir.}         \\ \hline
Video (UDP)              & 4        & 2           & 10    &  1400      & \gls{ul}                    \\ \hline
Sensor (UDP)             & 6        & 2           & 1.6   &  500      & \gls{ul}                   \\ \hline
Smartphone (TCP)  & 25     & {[}25 , 75{]} & X & 1440 (ACK 40) & \gls{ul} {+} ACKs in DL           \\ \hline
Smartphone (TCP) & 125    & {[}5 , 95{]}  & X & 1440 (ACK 40)& \gls{dl} {+} ACKs in UL         \\ \hline
\end{tabular}
\caption{Application settings, if a setting does not apply it is marked with an ``X''}
\label{table:app_e2e}
\end{table}

\begin{table}[!t]
\footnotesize
\centering
\begin{tabular}{|c|c|}
\hline
\textit{Parameter}                & \textit{Value}         \\ \hline
Channel Model            & 3GPP          \\ \hline
Channel Condition        & Line-Of-Sight \\ \hline
Channel bandwidth        & 100 MHz       \\ \hline
Channel central freq.    & 28 GHz        \\ \hline
Scenario                 & Urban (UMa)   \\ \hline
Shadowing                & false         \\ \hline
Beam Angle Step          & 10 degrees    \\ \hline
Beamforming Method       & Beam Search   \\ \hline
Modulation Coding Scheme & Adaptive      \\ \hline
Ctrl/Data encode latency & 2 slots       \\ \hline
Radio Scheduler          & Round-Robin   \\ \hline
\end{tabular}
\caption{Relevant simulation parameters}
\label{table:sim_param}
\end{table}

For the traffic types, each of the video and sensor nodes has one UL UDP flow towards a remote node on the Internet. These flows are fixed-rate flows: we have a continuous transmission of 10 Mb/s for the video nodes, to simulate a 720p24 HD video, and the sensors transmit a payload of 500 bytes each 2.5 ms, that gives a rate of 1.6 Mb/s. Table~\ref{table:app_e2e} summarizes the UDP flow characteristics. For the smartphones, we use TCP as the transmission protocol, with the state-of-the-art ns-3 implementation~\cite{tcp_nat,SACK}. Each \gls{ue} has to download five times a 5 MB file (so the downloads count as five different flows) and to upload one file of 15 MB. These flows start at different times: the upload can start at a random time between the 25th and the 75th simulation seconds, while each download can start between the 5th and the 95th simulation seconds. Table~\ref{table:app_e2e} summarizes the details.

\medskip\textbf{Simulations campaign.} We compare \gls{nr} numerologies, from $\mu=0$ to $\mu=4$, and analyze the TCP goodput (the average rate at which the receiver application gets the data) and the UDP one-way delay (the average latency of each UDP packet from source to destination). Other relevant parameters for the simulations are reported in Table~\ref{table:sim_param}. For each $\mu$, we have performed multiple sets of simulations in the ns-3 network simulator to obtain data statistics. We have done the experiments using different decoding latencies, represented by the parameter \emph{decodingLatency}.

We consider four values for the decoding latency setting: 1) the ideal condition, in which the signal takes no time to reach the \gls{mac} layer (0 ms case); 2) a fixed value of 0.1 ms, representing high-speed decoding; 3) a fixed value of 0.5 ms, as in literature~\cite{7869608}; and 4) a slot-dependent latency value that is equal to two times the slot length (so that it varies accordingly with the numerology). Inside a single simulation, we average the flow performance of each class (video, sensor, TCP download, TCP upload) by using a geometric mean.

To obtain statistical significance, we repeated the same simulations using five different random seeds. In this way node positions, flow start times, and many other factors result randomized. Then, we use the geometric mean to average the result of the same traffic class with different seeds.

\begin{figure}[!t]
	\centering
	\includegraphics[width=0.8\columnwidth]{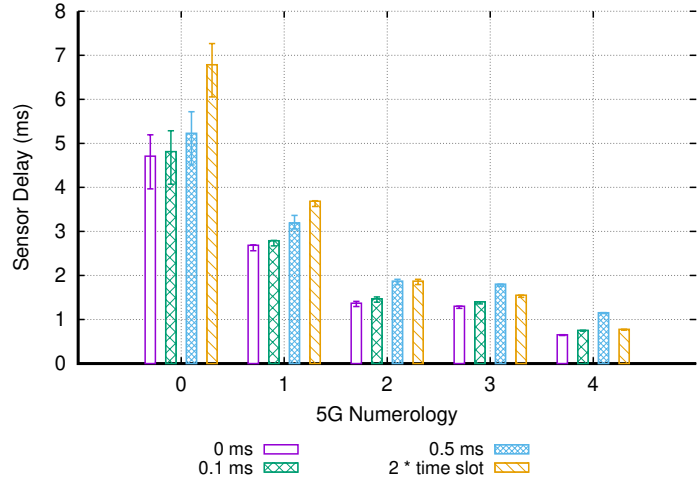}
	\caption{Sensor delay. Top whisker represents the maximum value, the bottom whisker represents the minimum value, and the box is at the 80th percentile.}
	\label{fig:sensor_delay}
\end{figure}

\medskip\textbf{Sensor and Video UDP Delay.} In Figure~\ref{fig:sensor_delay} we can see the latency performance of the sensor flows. In the first two numerologies, the worst performance is achieved by the delay configuration that is tied to the slot length. The explanation naturally follows if we keep in consideration that, in these numerologies, the slot length is much more than the fixed values we are considering. Instead, when the slot length is reduced, the performance starts to equal the fixed delays (the perfect example is represented by the equality, for $\mu{=}2$, of the last two cases: in fact, the slot length is equal to 0.25 ms, exactly half of the fixed delay of 0.5 ms). The best performance is offered by the ideal case of 0 ms decoding latency. In absolute values, increasing the decoding latency from 0 ms to 0.1 ms adds approximately 0.1 ms to the latency performance. The linear increase also applies when passing from 0.1 ms to 0.5 ms: that difference is added, almost without change, in the end-to-end delay value. These observations allow us to conclude that \emph{the analyzed fixed delays in the decoding impact the overall latency linearly}, without affecting other phases.

\begin{figure}[!t]
	\centering
	\includegraphics[width=0.8\columnwidth]{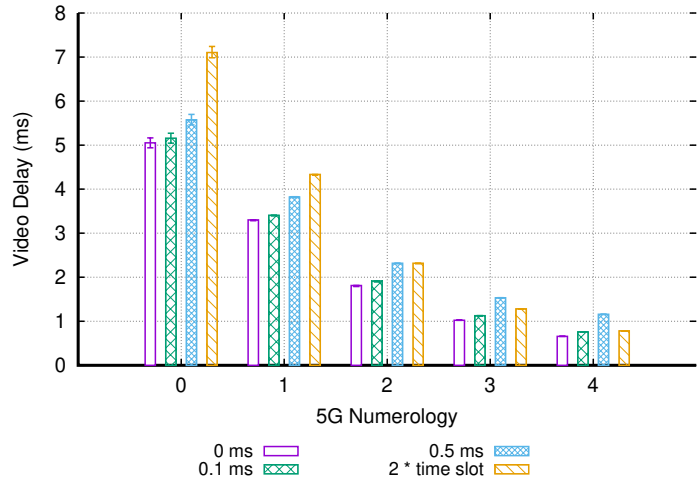}
	\caption{Video delay. Top whisker represents the maximum value, the bottom whisker represents the minimum value, and the box is at the 80th percentile.}
	\label{fig:video_delay}
\end{figure}

For the latency performance of the Video flows, we can refer to Figure~\ref{fig:video_delay}. Here we observe a similar trend to that shown by the sensor delay, but with slightly higher values.

\medskip\textbf{UDP Remarks.} Looking at the fixed-value case, with decoding latency of 0 ms, 0.1 ms, and 0.5 ms, we can see that there is a strange case in which an increased numerology corresponds to an increase in latency (or at least, not in a latency reduction). The increase happens in the sensor flows when passing from $\mu {=} 2 $ to $\mu {=} 3 $. How is it possible that half the slot time corresponds to more latency experienced by a single packet? The reason lies in the \gls{sr} mechanism. Before doing an \gls{ul} transmission, it is necessary to have the \gls{ul} Grant from the \gls{gnb}. A grant comes from an explicit \gls{sr}, or following a \gls{bsr} message sent along with user data in a previously granted space. If a data packet meets an empty \gls{rlc} buffer, the \gls{ue} is forced to send the \gls{sr} message to get an \gls{ul} grant from the gNB \gls{mac} scheduler. On the other side, if the \gls{rlc} buffer already contains data at the time the packet arrives, it is very likely that the \gls{ue} sent earlier the \gls{sr}, and all the upcoming data (until the buffer will be emptied) will be sent in grants that come automatically after the \gls{bsr}s.

The data arrival rate in the \gls{rlc} buffers, together with the transmission and the processing time, determines if the \gls{ul} flow needs a \gls{sr}, or it can rely on the \gls{bsr}, to continue the \gls{ul} transmission. The data arrival rate is fixed in all the experiments, while the transmission and processing time change with the numerology. The two components generate a synergy for which it is necessary to send a \gls{sr} to re-start the data flow. This generates the unfortunate event in which for the sensors, in numerology 2 the number of \gls{sr} is lower than in the numerology 3. Even if the slot time is lower, the overhead for the increased number of \gls{sr} is reflected in the latency value plotted in Figure~\ref{fig:sensor_delay} ($\mu {=} 2$). We do not see the same effect for video since the phenomenon is correlated to the inter-packet arrival time in the \gls{rlc} buffers, that are different between sensor and video flows.

\begin{figure}[!t]
	\centering
	\includegraphics[width=0.8\columnwidth]{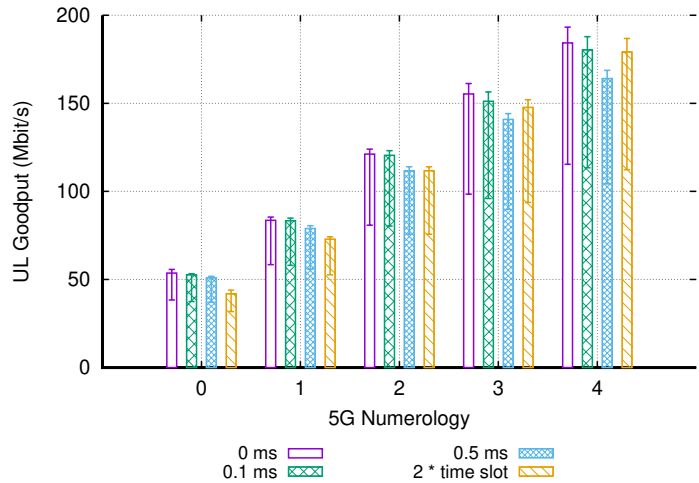}
	\caption{TCP Upload goodput. Top whisker represents the maximum value, the bottom whisker represents the minimum value, and the box is at the 80th percentile.}
	\label{fig:ul_th}
\end{figure}

\medskip\textbf{TCP Upload Goodput.} We start the analysis by looking at the goodput for the TCP upload flows, in Figure~\ref{fig:ul_th}. With higher numerologies, the slot time reduces, allowing to have smaller \gls{rtt}. Therefore, the goodput increases, reaching almost 200 Mb/s for the numerology 4. We observe that normally best results are obtained for the case of 0 ms processing delay. Another important thing is that the processing delay dependant on the slot length offers the worst performance in the lowest numerology (0 and 1), but starts to recover (and eventually in the last numerology outperforms the others) with the reduction of the slot time itself, due to the increasing numerology.

\begin{figure}[!t]
	\centering
	\includegraphics[width=0.8\columnwidth]{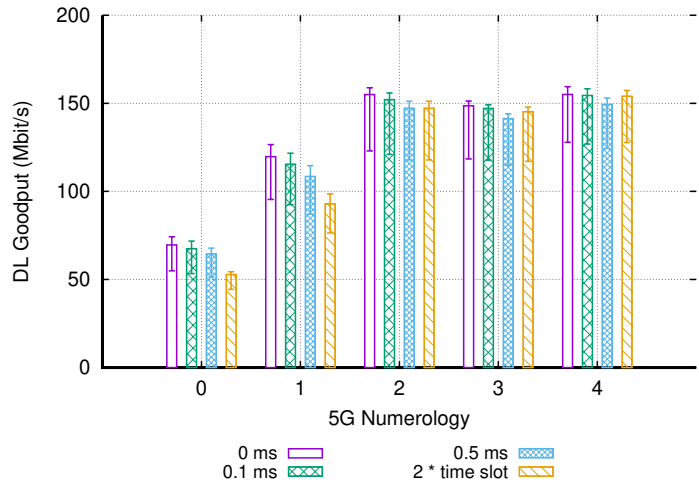}
	\caption{TCP Download goodput. Top whisker represents the maximum value, the bottom whisker represents the minimum value, and the box is at the 80th percentile.}
	\label{fig:dl_th}
\end{figure}

\medskip\textbf{TCP Download Goodput.} In Figure~\ref{fig:dl_th} we can analyze the performance of the TCP Download flows. In absolute values, the downloads have a slightly higher performance compared to uploads. In particular, comparing the upload and the download goodput in the same numerology, it is easy to see that \emph{the download goodput is almost 10 Mb/s higher than the upload goodput}. This difference is due to the absence, in the \gls{dl}, of the \gls{sr}/\gls{ul} Grant control messaging. When the data arrives in the buffers of the \gls{gnb}, it will take scheduling decisions almost immediately. When the data is waiting in the \gls{ue}'s buffer, instead, the permission to transmit is not immediate, but has to be granted by the \gls{gnb}, involving a signaling exchange that, albeit slightly, increases the round trip time and therefore reduces the TCP goodput. The trend of the goodput while increasing the numerology follows what we have seen in the TCP Upload, but stops after numerology 2. This is because our scheduler prioritizes  the \gls{ul} flows by assigning more usable symbols. So, at one point, the download flows are reaching their cap assigned by scheduler.

\section{Future Work}
\label{sec:future}
In this section, we describe our roadmap and future development plans regarding different layers of the protocol stack, according to NR specifications.

\subsection{PHY}

\subsubsection{Mini-slots}

NR defines mini-slots composed of 2 OFDM symbols up to the slot length - 1 in any band~\cite{R1-1708121},~\cite[Sect. 8.1]{TR38912}, and of 1 symbol, at least above 6 GHz. Although we already support flexible transmission duration thanks to the variable \gls{tti} concept, true mini-slots also include the \gls{pdcch}. In the current implementation, we transmit \gls{pdcch} only in the first symbol of a slot regardless of the \gls{tti} number in such slot. So, this will be extended by including the \gls{pdcch} inside the first symbol of the \gls{tti} allocation.

\subsubsection{PHY layer abstraction}
The current version of the simulator imports the PHY layer abstraction of LTE. One of the important changes of NR with respect to LTE (which used Turbo Codes) is the adoption of Polar Codes for control channels and \gls{ldpc} coding for data channels. Turbo Codes and \gls{ldpc} are shown to have similar performances for large packet sizes, however, the differences lie in the implementation complexity and when packet sizes are small. In this regard, we are currently working to include the specific NR PHY layer abstraction through a proper \gls{l2sm} and error model of NR. We are following the same approach like that of the ns-3 \gls{lte} module, in which the multi-carrier compression metrics are combined with link-level performance curves matching, to obtain lookup tables of \gls{bler} versus \gls{sinr}. The main objective is to predict the \gls{tbler} at \gls{mac} layer~\cite{mezzavilla:12}. In addition to the error model that has to be abstracted from a link-level simulator, the \gls{cb} sizes and the number of \gls{cb}s that map to a \gls{tb} need to be updated as per~\cite[Sect. 5.2.2]{TS38212}.

\subsubsection{256-QAM} NR supports 256-QAM modulation, which is still not available in the NR module. For that, we need to update the \gls{mcs}s as per~\cite[Table 5.2.2.1-2]{TS38214}, include the \gls{mi} mapping for 256-QAM, for which the mapping in~\cite{yang:14} can be used, and add new lookup tables of \gls{bler} versus \gls{sinr} for these new MCSs. We are working on this, which goes in parallel with the PHY layer abstraction, and will be released soon.

\subsubsection{Spatial user multiplexing and MIMO} Currently, we only support single-beam capability at a time, multiplexing of UEs in frequency/time domain, and single stream per UE (i.e., we have beamforming gain supported, but not spatial multiplexing gain). To support transmission/reception to/from multiple UEs simultaneously at the same time/frequency resources, we should extend the interference model and add power distribution per UE. To support the transmission of multiple streams per UE, the PHY abstraction model could be extended to support multiple streams per UE, for which also the precoders would need to be updated and redesigned.

\subsubsection{Frequency Division Duplexing} Currently the simulator supports dynamic TDD, which is the interesting choice for higher frequency ranges and is the approach currently mainly considered for deployment due to regulatory requirements, especially in the United States. However, support for FDD will be needed to simulate future deployments, as the regulation will progress.

\subsection{MAC}
\subsubsection{UL grant-free scheme} NR has introduced a contention-based (non-scheduled) access scheme for URLLC, named UL grant-free or autonomous UL~\cite[Sect. 8.1.2.1]{TR38802}~\cite{singh:18}. In UL grant-free, the UE is allowed, upon activation, to transmit UL data on resources devoted to contention-based access without a UL grant. Therefore, the UL grant-free scheme eliminates the handshake of SR, BSR, and UL grant, as compared to the UL grant-based access shown in Figure~\ref{fig_UL}. The resource allocation for UL grant-free in NR is as follows: time-domain resources for UL grant-free are configured by RRC signaling, and then the activation/deactivation is done through the DCI in PDCCH, which indicates to the UE the RBs and MCS to use if UE wants to access such resources. Its main problem is that, as it is non-scheduled, collisions may arise. Accordingly, to introduce this model in the ns-3 NR simulator we would need to (i) define resources for UL grant-free access, to be configured by RRC, and (ii) to include an error model for the collisions therein.

\subsubsection{Bearer prioritization} Currently, the scheduler is not prioritizing flows based on their bearer. In other words, it does not take into account guaranteed bit rate or latency deadlines when making decisions. A crucial step will be to insert a general policy to ensure such constraints to the active flows.

\subsubsection{Punctured scheduling} To efficiently multiplex eMBB and URLLC traffics, NR has defined procedures to enable punctured scheduling in \gls{dl}~\cite{klaus:18}. This is useful in case of a sudden need for resources for URLLC traffic that has strict latency requirements. URLLC latency targets can be met by puncturing the resources already allocated to eMBB traffic and indicating so to those UEs through an indicator of preemption. To support punctured scheduling in the NR simulator we should: (i) allow the scheduler to work symbol-by-symbol, instead of in a per-slot basis, when there is a new URLLC packet arrival in the middle of the slot, (ii) include an indication of preemption in DL control channels to indicate preemption of eMBB data, and (iii) redesign procedures for eMBB flows to ignore scheduling assignments and avoid decoding, as well as the subsequent HARQ-ACK feedback generation.

\subsection{Upper layers extensions}
Currently \gls{rrc}, \gls{rlc}, \gls{pdcp} layers are relying on LTE  implementation. Different simplifications have been proposed for RLC and PDCP, in order to facilitate the targeted latency reduction that NR should support. RRC needs also different extensions already in its original implementation, since it was mainly designed for operation in CONNECTED mode. As a result of that, updates and extensions will be considered for inclusion.

\subsection{SDAP}
In the new \gls{qos} framework for the 5G network, there is a new layer above \gls{pdcp}. Its name is \gls{sdap}, and its role is to map distinct \gls{qos} flows into data radio bearers. In the simulator, we miss an SDAP entity that receives \gls{sdu} from upper layers and sends SDAP \gls{pdu} to its peer SDAP entity via lower layers. These entities should be able to mark the \gls{qos} flows appropriately, and it should be possible to configure the mapping between flow and data radio bearer, through a static or a dynamic configuration via RRC.

\subsection{Core Network}
In 5G, the core network will support the separation of control and user plane, and each service will be provided as a network function, creating an overall service-based architecture. We are incorporating changes from ns-3 LTE module, such as the EPC functional split between \gls{pgw} and \gls{sgw} that has been recently included in LTE ns-3, but we should prepare a compatible API to offer functionalities to external users.

\subsection{Operation in unlicensed bands}
Differently from LTE, NR includes native support for operation in unlicensed bands. A work item is currently ongoing to define the operation in the band below 7 GHz, and for Release 17 extensions are expected for mmWave bands. The operation will include Listen Before Talk functionalities, able to facilitate coexistence in the same band with technologies like WiFi and WiGig. Thanks to the multi-technology characteristic of ns-3, we will be able to propose such an extension in our module.

\section{Conclusion}
\label{sec:conclusions}
In this paper, we have presented a complete overview of a novel full-stack NR simulator. In particular, we have discussed the additions and modifications that we have done to the mmWave simulation tool developed on top of ns-3. The objective was to support an end-to-end simulation of NR networks, through an up-to-date and standard-compliant platform. The work has been validated and calibrated in different indoor scenarios, as compared to other proprietary simulators that follow similar purposes inside 3GPP, and by following 3GPP recommendations. In addition, as an example of the potentiality of the tool, we have created a complex end-to-end simulation campaign to assess the impact of different NR numerologies on the overall E2E latency of different devices. We  have  studied  parameter  sensitivities and shown that  when  considering a full protocol stack and high fidelity models, unexpected behaviors are observed, which hardly could be highlighted with other types of simulators. We have concluded the discussion with the presentation of the roadmap that we are currently following, and wish to follow, in line with current and future 3GPP developments.

\section*{Acknowledgment}
This work was partially funded by Spanish MINECO grant
TEC2017-88373-R (5G-REFINE) and Generalitat de Catalunya grant 2017 SGR 1195. Also, it was supported by InterDigital Communications, Inc.

\section*{References}
\bibliography{main}

\end{document}